\documentclass[12pt,a4paper]{article}
\usepackage[latin1]{inputenc}
\usepackage{graphicx,hyperref,lipsum,titling,url,tikz,multirow,algcompatible,subfig}
\usepackage[ruled]{algorithm2e}
\usepackage[cm]{fullpage}
\DeclareGraphicsExtensions{.pdf,.eps,.png,.jpg}
\newcommand{\zg}{``}
\newcommand{\zd}{''}
\author{Pascal Cotret $^\alpha$, Guy Gogniat $^\beta$, Martha Johanna Sepúlveda Flórez $^{\beta,\gamma}$\\
\small{$^\alpha$ IETR/SCEE, CentraleSupélec - FRANCE}\\ \small{\href{mailto:pascal.cotret@centralesupelec.fr}{pascal.cotret@centralesupelec.fr}}\\
\small{$^\beta$ Lab-STICC Laboratory, University of South Brittany - FRANCE}\\ \small{\href{mailto:guy.gogniat@univ-ubs.fr}{guy.gogniat@univ-ubs.fr}}\\
\small{$^{\beta,\gamma}$ Institute for Security in Information Technology, Technical University of Munich - GERMANY}\\
\small{\href{mailto:johanna.sepulveda@tum.de}{johanna.sepulveda@tum.de}}}
\title{Protection of heterogeneous architectures on FPGAs: An approach based on hardware firewalls}
\date{}

\pretitle{\begin{center}\vspace{-2cm}\small{(pre-print version of a manuscript published in Elsevier Microprocessors and Microsystems)}\\~\\~\\\Large}
\posttitle{\end{center}}
\begin{document}
\maketitle
\vspace{-1.5cm}
\section*{Abstract}
\noindent Embedded systems are parts of our daily life and used in many fields. They can be found in smartphones or in modern cars including GPS, light/rain sensors and other electronic assistance mechanisms. These systems may handle sensitive data (such as credit card numbers, critical information about the host system and so on) which must be protected against external attacks as these data may be transmitted through a communication link where attackers can connect to extract sensitive information or inject malicious code within the system. This work presents an approach to protect communications in multiprocessor architectures. This approach is based on hardware security enhancements acting as firewalls. These firewalls filter all data going through
the system communication bus and an additional flexible cryptographic block aims to protect external memory from attacks. Benefits of our approach are demonstrated using a case study and some custom software applications implemented in a Field-Programmable Gate Array (FPGA). Firewalls implemented in the target architecture allow getting a low-latency security layer with flexible cryptographic features. To illustrate the benefit of such a solution, implementations are discussed for different MPSoCs implemented on Xilinx Virtex-6 FPGAs. Results demonstrate a reduction up to 33\% in terms of latency overhead compared to existing efforts.
\section{Introduction}%
For many years, embedded systems are used in our daily life: we found them in electronic devices, automotive applications, telecommunications systems and so on. When designing such systems, several issues have to be taken into account and one of the major concerns is about security. Since the late 90s, security has become a key point in the development of embedded systems \cite{Kocher04}. The number of weaknesses is in constant progress and electronic devices have to process data with various security requirements. According to \cite{Ravi04}, security criteria are communications security, storage security, inputs/outputs security and users authentication. This work focuses on the two first criteria (communications and storage).

First of all, this work considers communication protection as a key point in embedded systems development as communications channels convey several data types (application codes, confidential data, cryptographic elements and so on) with various needs in terms of security: confidential data must not be revealed to an unauthenticated user while application may be accessible through a specific interface (for instance, for development purposes). Then, this work also takes care of data storage security: memory elements are another critical entry point for attackers as they potentially contain plaintext data.

This work is organized as follows. Section \ref{sec_context} presents related works and our constraints regarding the architecture. Sections \ref{sec_static} and \ref{sec_dynamic} describe our solution in a static and dynamic approach. Section \ref{sec_res} gives an analysis in terms of security and provides implementation results in comparison with other approaches.

\section{Scientific context}\label{sec_context}
\subsection{Related works}
Several studies dealing with security in embedded systems have been published \cite{Kocher04,Ravi04}. Security mechanisms can be implemented in two ways: hardware blocks or software functions. Software solutions are generally slower, in terms of latency, than a pure hardware-implemented security solution. Furthermore software solutions are generally more easily compromised than hardware countermeasures. In this section, several works about memory protection are presented. Then, regarding internal transactions protection, an overview of the main solutions is proposed.

\subsubsection{Memory protection approaches}
In order to provide countermeasures against the threat model defined in Section \ref{subsec_thm}, a key point is to address memory protection. An obvious solution is to implement cryptographic features for memory confidentiality and integrity. XOM \cite{Lie03} is a solution mixing confidentiality and integrity for systems where the external memory can be tampered. The implementation requires adding hardware modules and modifying the processor structure. Using such a solution, performances are quite spoiled as authors \cite{Lie03} announce a 50\% loss. AEGIS architecture \cite{Suh07} is another approach based on a security-enhanced processor embedding confidentiality and integrity functions. Depending on processor configuration (cache size), memory slowdown is between 3.8\% and 130\%. Bossuet et al. \cite{Bossuet13} made an in-depth comparison of existing cryptographic processors where some of them, such as HCrypt, were implemented on FPGAs. Some of these processors are efficient but do not cover our threat model defined in a further section. In \cite{Mahar06} authors describe a solution (called SecSoft) to protect software with a hardware Encryption Management Unit (EMU). This work proposes a latency analysis of several modes (block/counter modes for encryption function and with/without encryption). Latency overhead on a ML301 platform goes from less than 10\% (block mode, unencrypted) up to 80\% (block mode, encrypted). This solution does not provide mechanisms targeting integrity. \cite{Elbaz06} proposes the PE-ICE solution to check integrity in parallel to encryption (i.e. confidentiality). The worst case implementation shows a performance loss of 20\% and for a confidentiality only implementation, a 4\% loss is given. Vaslin et al. \cite{Vaslin08} proposes a confidentiality and integrity hardware block based on AES for confidentiality and cyclic redundancy check for integrity, performance loss is about 13-14\%. In \cite{Crenne13}, authors extended this work by using the AES-GCM algorithm (this option is also used in this work), integrity is done by a low latency function \cite{Crenne11a}. Other methods such as hash trees and formal verification \cite{Haifeng14,Wiersema14} are use to protect memory contents.

Another solution is to use the built-in MMU (Memory Management Unit) available in some processors. This work is implemented on a Xilinx FPGA where the softcore Microblaze is provided as a general purpose processor. Microblaze MMU \cite{Xilinx12a} provides a simple access control allowing to get read-only or full-access memory pages in the system, this control can be disabled in a configuration register.

All these works propose solutions to provide encryption methods for external memories protection with different performance versus security tradeoff. In order to protect the target system from attacks on the external memory, a trivial solution consists in building a fully-protected external memory unit. Unfortunately, in this case, each memory access implies a ciphering/deciphering latency penalty. Thus, such an approach is strongly penalizing regarding the overall latency overhead of an application. To mitigate this point, our work proposes to implement cryptographic features only on specific memory pages, defined by application requirements, avoiding such a systematic latency penalty. Therefore, some pages are still not protected, that is the reason why internal traffic protection and/or monitoring must be also addressed in the context of embedded systems security.

\subsubsection{Bus and NoC-based security methods}
\begin{figure}[htbp]
	\centering
\includegraphics[width=.75\textwidth]{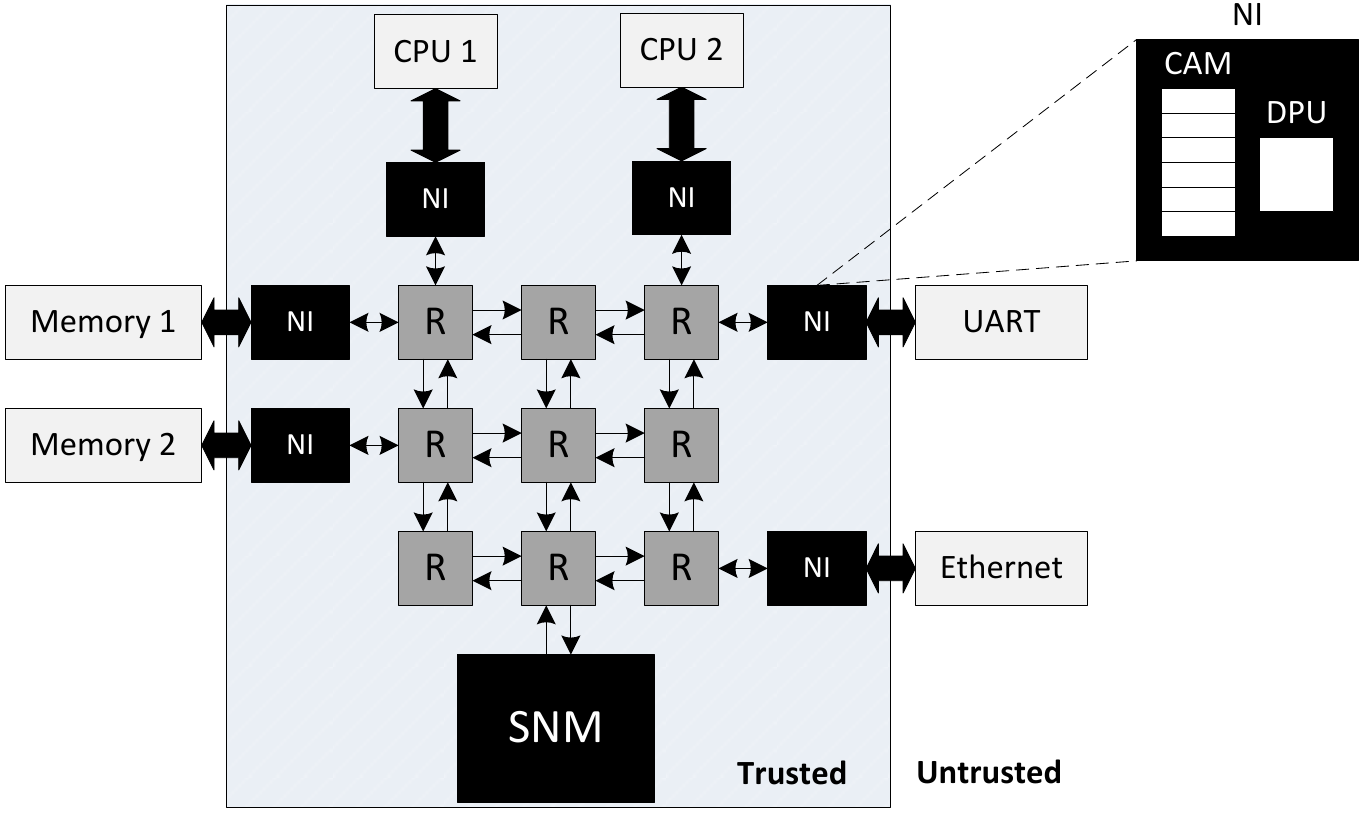}
	\caption{Fiorin's approach \cite{Fiorin08b,Fiorin07}}
	\label{fig1}
\end{figure}

Regarding large scale systems with NoC-based communication architecture, \cite{Diguet07} proposes a solution where security controls are done in each network interface in a distributed manner. In this case, a security manager unit gathers individual interfaces information and performs countermeasures and security updates (done through dynamic partial reconfiguration). This method takes into account processor facing denial of service attacks but does not offer ciphering function (however, authentication and integrity features are available). Fiorin et al. \cite{Fiorin07,Fiorin08a,Fiorin08b} propose an alternative to this approach providing ?security sensors? inside network interfaces to refine controls (NI in Figure \ref{fig1}). These sensors are able to block incoming malicious data when parameters are not proven. These parameters are stored in a trusted CAM (Context-Addressable Memory). Finally, a SNM (Security Network Manager) gathers information from NIs to detect potential collisions and errors in data traffic. Unlike \cite{Diguet07}, security mechanisms can be updated without partial reconfiguration of the FPGA chip, update is done using memory rewriting. \cite{Diguet07} and \cite{Fiorin07,Fiorin08a,Fiorin08b} do not offer cryptographic features to cipher data transmitted in the communication network.
~\\
The solution proposed by Fiorin et al. is based on a secured Network Interface (NI) with a DPU
mechanism (Data Protection Unit, see Figure \ref{fig1}). This mechanism allows or not a transaction according to parameters stored in individual trusted CAM memories. \cite{Fiorin07} requirements aim to cover a threat model with denial of service attacks but cannot protect systems against IP modifications performed by an attacker. This distributed approach has a low-latency and also presents some robustness. Controls are done in each interface, even if one of them is corrupted, other interfaces should still continue to work. Regarding NoC-based security solutions, we can also cite \cite{Sepulveda12a,Sepulveda12b}. Structure proposed by Sepulveda et al. is based on a hierarchical NoC with low NoCs (each low NoC is a sub-network having a single security policy) and a single high NoC acting as a global security manager (connections with each low NoC). Regarding NoC-based security, LeMay et al. \cite{LeMay2015} propose a mechanism to detect abnormal behaviors in a NoC protocol when a malicious IP is inserted in the system architecture: the solution based on AXI signals has an area overhead up to 23\%.

Then, for bus-based MPSoCs, the main contribution was published by Coburn et al. \cite{Coburn05}. This approach is similar to \cite{Fiorin07,Fiorin08a,Fiorin08b} as it is based on security-enhanced network interfaces called SEI (\textit{Security Enforcement Interface}) but in a centralized approach. The main drawback of this solution is that security information is sent to a global security manager which is the only component able to perform controls. Therefore, latency overhead is increased (see Section \ref{sec_res}). This solution does not offer security updates or cryptographic features. Coburn et al. approach suggests to centralize all the controls in a single module; therefore, as soon as this module is corrupted, system security is compromised.

Compared to these efforts, our approach provides a distributed solution with update mechanisms. We are able to dynamically adapt the security policies based on the instantaneous threats. We also provide some cryptography mechanisms in order to fully protect the system. We propose a low-latency solution in order to reduce the performance penalty due to security.

\subsubsection{Other methods}
Other solutions propose a physical separation of components in order to define secured and non-secured areas. For instance, \cite{Huffmire08} use \textit{moats} during place-and-route, routing is forbidden in these areas, it allows isolating specific areas of the FPGA. Other contributions propose to combine hardware and software elements to provide security. \cite{Plouviez11,Porquet2010} focus on virtualization, an hardware mechanism associated with software services that manages several software tasks in a secured manner on an heterogeneous architecture. Another solution is proposed by ARM, Trustzone \cite{Zu08}. It offers the possibility to have two areas with different privileges. A hardware module is in charge of monitoring data, communications between these two areas are secured if the parameters of the architecture are set to secure. However, virtualization-based solutions are not adapted to our context: for small-scale MPSoCs, it is assumed that hardware-based solutions are desirable. Furthermore, this work aims to provide a solution with the lowest impact on existing OS or architecture: as virtualization adds some mechanisms to both software and hardware layers, these solutions are not considered in our further comparison. Our contribution could be extended with mechanisms as provided by \cite{Huffmire08}. Our approach goes into the same direction as works developed by \cite{Plouviez11,Porquet2010} and \cite{Zu08}, but we propose a more comprehensive solution providing cryptography, filtering and monitoring.

\subsection{Platform characterization and threat model}\label{subsec_thm}
This work implies a set of constraints about architectures where our approach can be applied:
\begin{itemize}
	\item FPGA chips are chosen as they allow short development times and system reconfigurability whereas ASICs cannot be reprogrammed.
	\item This work is done in a context of a project where applications do not require a large number of IPs (small-scale up to medium-scale architectures). Therefore, it is assumed that a single communication bus is able to manage all case studies targeting MPSoCs.
	\item Regarding communication protocols, AXI protocol from ARM \cite{Arm12} is chosen. In fact, this protocol is the standard in latest Xilinx development tools and should provide a compatibility with other ARM-based technologies in the future such as Zynq circuits (embedding an FPGA and a Cortex-A9 processor \cite{Xilinx12b}) or Armadeus \cite{Armadeus14}.
\end{itemize}
This work is based on a specific threat model for FPGA-based MPSoC architectures. At a high level of abstraction, it is assumed that the FPGA itself is a trusted component (i.e. attackers cannot directly access it). On the opposite, both external bus and external memory are considered as untrusted areas. As a consequence, we can define a range of attack scenarios that must be covered by our approach. On the one hand, following scenarios are taken into account (non exhaustive list):
\begin{itemize}
	\item Attack A1. External memory is compromised by replacement or modification. In this scenario, it is considered that the attacker is able to physically replace the external memory by its own chip. The attacker is also able to modify its contents by any means: accesses can be logical (access with another processor) or even physical (for instance, if memory contents are modified by overheating or exposure to an electromagnetic field).
	\item Attack A2. Snooping on-chip to off-chip bus through an external probe. The attacker is able to access the physical bus between the FPGA chip and the external memory. The attacker can read data on the bus or inject malicious data.
	\item Attack A3. Internal snooping on on-chip busses for data values by an untrusted IP included in the SoC. It is considered that the attacker is able to implement an IP (hardcore or softcore) aiming to add data to the on-chip traffic of the system. These data can be both packets with valid address or denial-of-service attempts. The attacker is also able to retrieve data from the bus for example to get private information.
\end{itemize}
On the other hand, following scenarios are not considered in our threat model:
\begin{itemize}
	\item Snooping on the on-chip busses through package modification or physical probe. This work considers that the chip cannot be physically modified after bitstream downloading.
	\item Attempting to measure the contention on the bus or cache-timing attacks.
	\item Other side-channel attacks: power analysis, electromagnetic analysis and so on.
\end{itemize}
These scenarios can be combined to construct complex attack scenarios. Two scenarios will be tested in Section \ref{subsec_secana} to compare this work with existing techniques. As countermeasures, this work proposes to focus on protecting and monitoring the internal traffic as well as protecting the external memory.

\subsection{Main contributions}
Contributions provided in this work aim to protect external memories and internal traffic in multiprocessor architectures implemented on FPGA circuits. Our approach provides the demonstration that a complete end-to-end solution is possible. Such an approach allows reducing the performance penalty due to cryptographic features as all communications are not encrypted while still guaranteing that unprotected code or data will not lead to an attack on the system thanks to internal traffic protection/monitoring. Our solution represents an interesting alternative compared to \cite{Fiorin07,Coburn05} with an additional security layer including cryptographic services.
\begin{table}[htbp]
	\centering
	\caption{Comparison of existing solutions}
	\label{tab1}\tiny{
		\begin{tabular}{|c|c|c|c|c|c|c|c|c|c|}
			\cline{3-10}
			\multicolumn{2}{l|}{} & {\bf \cite{Crenne13}} & {\bf \cite{Xilinx12a}} & {\bf \cite{Lie03,Suh07}} & {\bf \cite{Diguet07}} & {\bf \cite{Fiorin08a}} & {\bf \cite{Sepulveda12b}} & {\bf \cite{Coburn05}} & {\bf This work} \\ \hline
			\multicolumn{2}{|c|}{{\bf \begin{tabular}[c]{@{}c@{}}Communication\\ technology\end{tabular}}} & N/A & N/A & N/A & \begin{tabular}[c]{@{}c@{}}NoC\\ (mesh)\end{tabular} & \begin{tabular}[c]{@{}c@{}}NoC\\ (mesh)\end{tabular} & \begin{tabular}[c]{@{}c@{}}NoC\\ (hierarchical)\end{tabular} & \begin{tabular}[c]{@{}c@{}}Bus\\ (AMBA)\end{tabular} & \begin{tabular}[c]{@{}c@{}}Bus\\ (AXI)\end{tabular} \\ \hline
			\multicolumn{2}{|c|}{{\bf Approach}} & \begin{tabular}[c]{@{}c@{}}Mem.\\ extension\end{tabular} & \begin{tabular}[c]{@{}c@{}}Proc.\\ extension\end{tabular} & \begin{tabular}[c]{@{}c@{}}Proc.\\ extension\end{tabular} & \begin{tabular}[c]{@{}c@{}}Centralized\\ IP\end{tabular} & \begin{tabular}[c]{@{}c@{}}Distributed\\ interfaces\end{tabular} & \begin{tabular}[c]{@{}c@{}}Distributed\\ IPs\end{tabular} & \begin{tabular}[c]{@{}c@{}}Centralized\\ IP\end{tabular} & \begin{tabular}[c]{@{}c@{}}Distributed\\ interfaces\end{tabular} \\ \hline
			\multicolumn{2}{|c|}{{\bf Security updates}} & No & No & No & No & Yes & No & No & Yes \\ \hline
			\multirow{5}{*}{{\bf \begin{tabular}[c]{@{}c@{}}Counter-\\ measures\end{tabular}}} & {\bf Confidentiality} & Yes & No & Yes & No & No & No & No & Yes \\ \cline{2-10} 
			& {\bf Integrity} & Yes & No & Yes & No & No & No & No & Yes \\ \cline{2-10} 
			& {\bf \begin{tabular}[c]{@{}c@{}}Memory\\ partitioning\end{tabular}} & Yes & Yes & No & No & Yes & Yes & Yes & Yes \\ \cline{2-10} 
			& {\bf \begin{tabular}[c]{@{}c@{}}Traffic\\ monitoring\end{tabular}} & No & No & No & Yes & Yes & Yes & Yes & Yes \\ \cline{2-10} 
			& {\bf \begin{tabular}[c]{@{}c@{}}Access\\ control (R/W)\end{tabular}} & No & Yes & No & Yes & Yes & Yes & Yes & Yes \\ \hline
			\multirow{3}{*}{{\bf \begin{tabular}[c]{@{}c@{}}Threat\\ model\\ coverage\end{tabular}}} & {\bf Attack A1} & Yes & No & Yes & No & No & No & No & Yes \\ \cline{2-10} 
			& {\bf Attack A2} & Yes & No & Yes & No & No & No & No & Yes \\ \cline{2-10} 
			& {\bf Attack A3} & No & No & No & Yes & Yes & Yes & Yes & Yes \\ \hline
		\end{tabular}}
	\end{table}
	~\\
	Table \ref{tab1} presents a summary of qualitative parameters targeted by our solution (which is basically described in Section \ref{sec_static}). This work enhances existing efforts by providing a solution for a comprehensive protection of internal and external communications:
	\begin{itemize}
		\item For traffic monitoring and protection, our approach is similar to \cite{Fiorin07} but it provides a more efficient solution for security updates in terms of memory consumption. This contribution is detailed in Section \ref{sec_res}.
		\item For memory countermeasures, our approach is similar to \cite{Mahar06,Elbaz06} but the AES-GCM cryptographic primitive is used as proposed by \cite{Crenne13}. Such a solution allows reducing the cryptography latency overhead while maintaining a strong level of protection. Compared to \cite{Mahar06,Elbaz06}, we rely on a more comprehensive security policy where developer can finely tune his/her cryptography needs.
	\end{itemize}
	Then, this work also presents a feedback feature for security mechanisms (Section \ref{sec_dynamic}): that is to say how attacks are detected and how the update manager takes care of these information. Finally, an evaluation in terms of implementation results and security analysis is given.
	
	\section{Static security for communications and memories}\label{sec_static}
\begin{figure}[htbp]
		\centering
		\includegraphics[width=.75\textwidth]{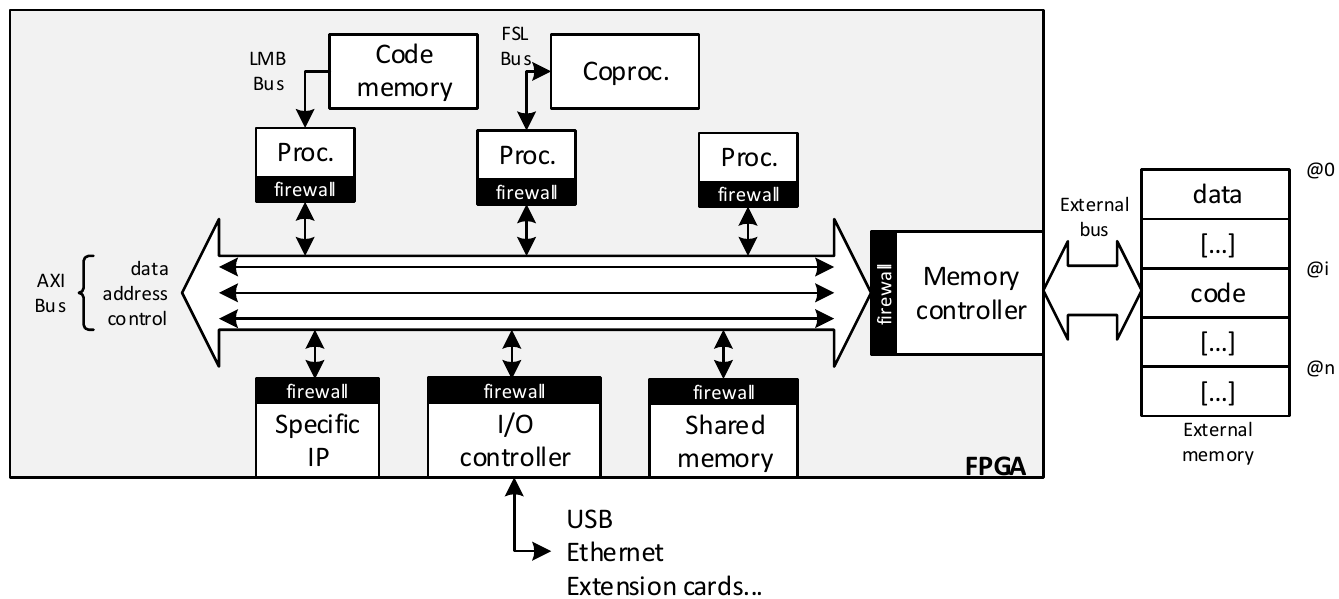}
		\caption{Generic MPSoC with hardware firewalls}
		\label{fig2}
	\end{figure}

	In order to secure MPSoCs, this work proposes an approach with hardware security components (also known as \zg firewalls\zd) implemented in the architecture. Figure \ref{fig2} shows an example of such a system.
	~\\
	A key point when dealing with security corresponds to the need of clearly setting up all parameters chosen to protect an MPSoC against a given threat model. Thus before discussing our approach, a set of rules (or security policies) is defined for each access in the system.
	
	\subsection{Security policy}
	Each security policy is defined by several parameters:
	\begin{itemize}
		\item The memory segment base/high addresses. An essential feature of firewalls is to block illegal accesses in terms of addresses. For instance, an illegal access to a memory section by a general purpose processor which is not allowed by the application specifications must be discarded (writing a confidential data 190 into memory, reading a section with cryptographic-related information\ldots).
		\item In order to avoid attacks such as \zg buffer overflow\zd, data format has to be analyzed for each transaction (both read and write operations).
		\item As systems contain an external memory, cryptographic primitives are used for data protection. According to the application requirements, some memory sections may not be ciphered if the contents are not critical.
	\end{itemize}
	As described in Figure \ref{fig2}, a generic MPSoC contains an external memory. Thus the most radical solution to guarantee security consists in protecting the whole memory in terms of confidentiality and authentication. In such a case, an attacker cannot decipher or modify its contents. Unfortunately, this solution has an important overhead in terms of latency (see Section \ref{sec_res}). An alternative solution consists in ciphering only the most critical memory sections. In this case, an attacker can read and write all the other plaintext memory
	sections. That is the reason why implementation of a cryptographic function only is not efficient enough to protect the target MPSoC against the defined threat model. Therefore, firewalls also aim to protect the system against these plaintext contents by monitoring the internal communications.
	
	\subsection{Local Firewall}
\begin{figure}[htbp]
		\centering
		\includegraphics[width=.75\textwidth]{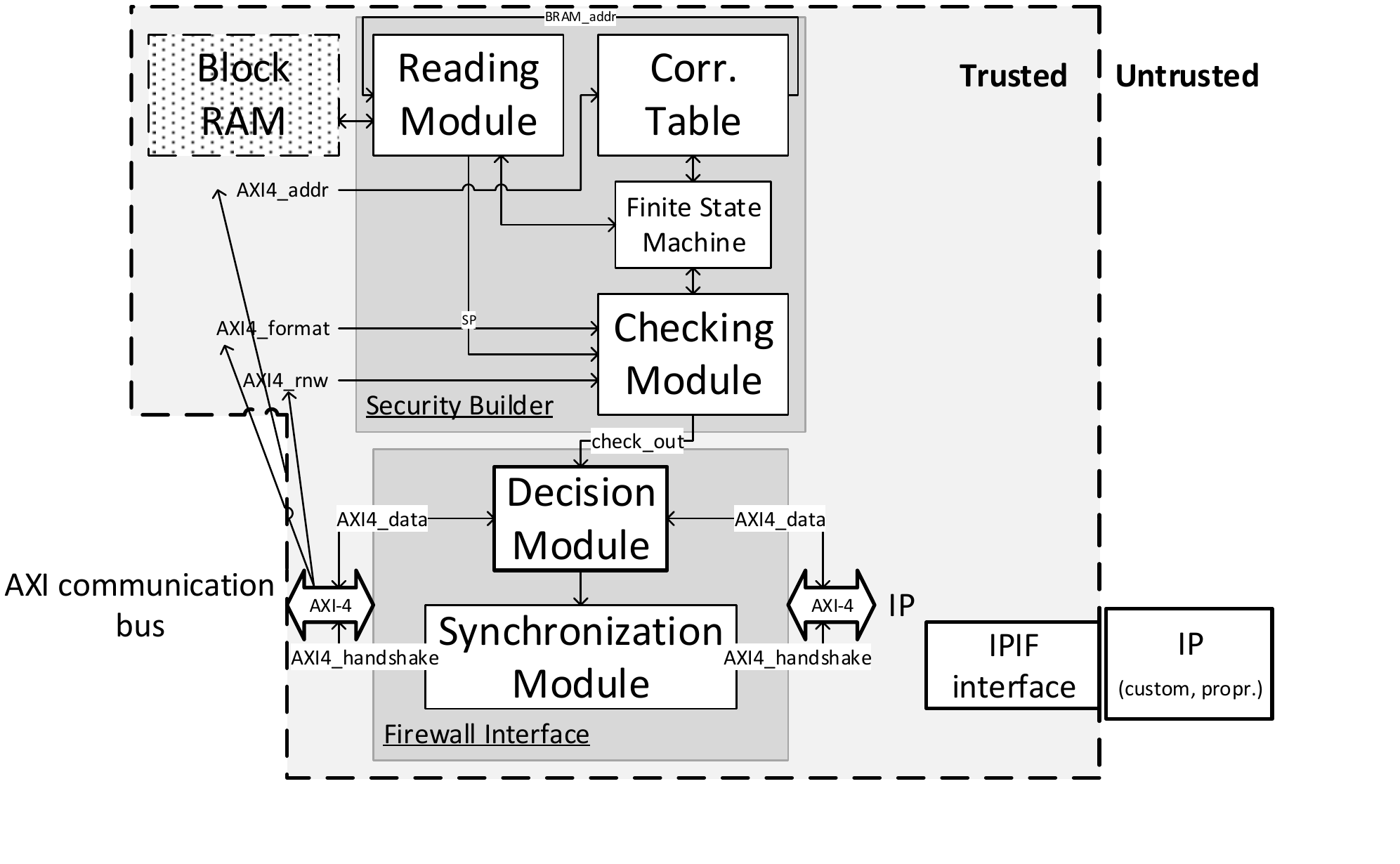}
		\caption{Structure of a Local Firewall}
		\label{fig3}
	\end{figure}

	Each firewall is composed of several blocks (\emph{Security Builder} and \emph{Firewall Interface}) connected as defined in Figure \ref{fig3}. The \emph{Security Builder} module is responsible for security policy management and the \emph{Firewall Interface} is a mandatory checkpoint towards the external world (that is to say, the system communication bus). This module acts as a gateway between the AXI-4 communication bus and the associated IP (it can be dedicated to an application, an I/O controller, a memory interface\ldots).
	~\\
	The whole firewall structure is considered as a trusted area (it is implemented in hardware in the FPGA chip), the only part to be untrusted is composed of the external bus and the external memory (components linked to the \emph{Cryptographic Firewall}).\newpage
	
	\subsubsection{Firewall Interface}
\begin{figure}[htbp]
		\centering
		\includegraphics[width=.75\textwidth]{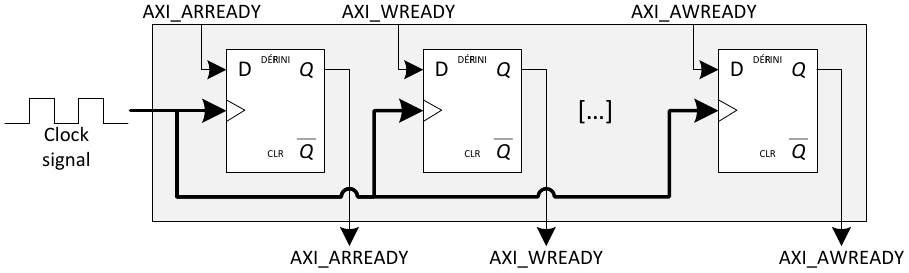}
		\caption{Structure of a Synchronization Module}
		\label{fig4}
	\end{figure}

	The \emph{Firewall Interface} module (\#1 in Figure \ref{fig3}) performs mainly two tasks:
	\begin{itemize}
		\item Once a data block has been analyzed and validated (its parameters match with the associated security policy), the \emph{Firewall Interface} transmits data to the transaction target (communication bus or IP); this operation is done by the \emph{Decision Module}.
		\item The Firewall Interface synchronizes the communication protocol signals such as handshake signals (\emph{AXI\_WVALID}, \emph{AXI\_RREADY}\ldots) or other control signals (\emph{AXI\_WSTRB}, \emph{AXI\_WLAST}\ldots) to keep valid transactions. Data block is synchronized with its control signals in order to avoid data loss and/or duplication. The sub-component \emph{Synchronization Module} is composed of a set of flip-flops (see Figure \ref{fig4}) where the clock port (at least its rising edge) is the acknowledgment signal \emph{check\_out} sent by the \emph{Security Builder} when all the data checking has been performed.
	\end{itemize}
	%
	As all the flip-flops are connected in parallel, the overall \emph{Firewall Interface} latency is of 2 cycles for each data block: 1 cycle for the \emph{Decision Module} and 1 cycle for the synchronization step.
	
	\subsubsection{Security Builder}
	The \emph{Security Builder} module is the main component within firewalls (\#2 in Figure \ref{fig3}). It is composed of 4 submodules:
	\begin{itemize}
		\item \textbf{Correspondence Table (CorrTable)}. Security policies are stored in a Block RAM memory considered as a trusted entity (upper left corner in Figure \ref{fig3}) and can be identified by an address. Internal structure of \emph{Correspondence Table} is detailed in Figure \ref{fig5}. Each policy is defined for a given physical address space (with lower and upper bounds), transcription towards policies addresses is done as defined in Algorithm \ref{alg1}.\newpage
		\begin{figure}[htbp]
			\centering
			\includegraphics[width=.75\textwidth]{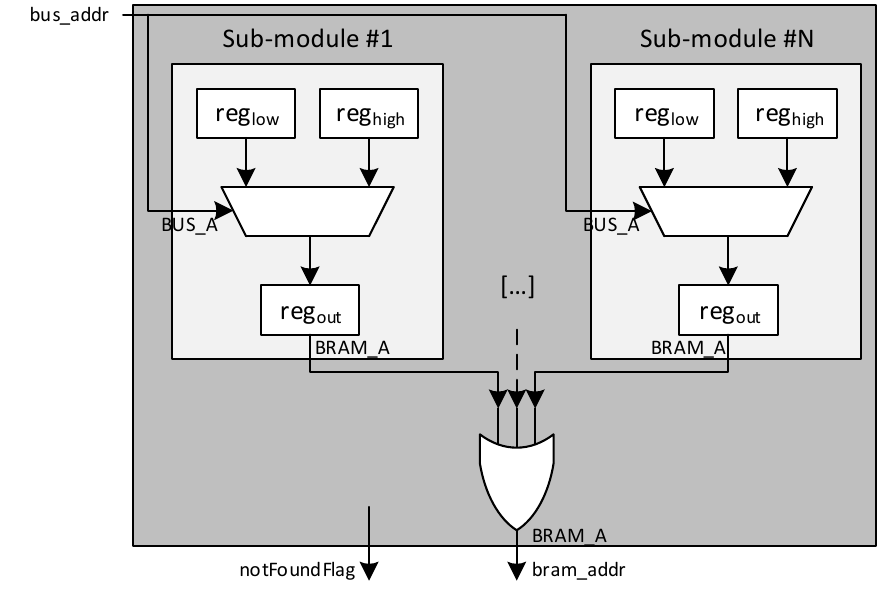}
			\caption{Implementation of a Correspondence Table}
			\label{fig5}
		\end{figure}
		Given that $N$ is the number of entries of the \emph{Correspondence Table} (i.e. the number of security policies embedded in the firewall), \emph{BRAM\_A} is initially equal to zero. If the bus address is contained in a given address space (defined with $reg_{low}$ and $reg_{high}$), \emph{BRAM\_A} is set to this value. Otherwise if \emph{BUS\_A} is outside the address space, a flag is set to high value (this flag reports an error and is treated by the global finite state machine embedded in each firewall).
		\begin{algorithm}[htbp]
			\caption{Correspondence Table loookup}
			\label{alg1}
			\begin{algorithmic}[1]
				\REQUIRE $n \leq N_{max}$ \COMMENT{$N_{max}$: \texttt{max number of entries}}
				\STATE $BUS\_A \leftarrow bus\_addr$ \COMMENT{\texttt{Get a copy of bus address}}
				\STATE $BRAM\_A \leftarrow 0x00000000$ \COMMENT{\texttt{Initialize output}}
				\FOR{$SUBMODULE=1$ to $N$}\COMMENT{\texttt{For each submodule}}
				\IF{$BUS\_A \in [reg_{low};reg_{high}[$} \COMMENT{\texttt{If submodule contains the bus address}}
				\STATE $BRAM\_A \leftarrow reg_{out}$ \COMMENT{\texttt{Write the location of security policy}}
				\ENDIF
				\ENDFOR
				\IF{$BRAM\_A=0$} \COMMENT{\texttt{If the output register is null}}
				\STATE $not\_found\_flag \leftarrow 1$ \COMMENT{\texttt{Set a warning flag !}}
				\ENDIF
				\STATE $bram\_addr \leftarrow BRAM\_A$ \COMMENT{\texttt{Export the output to other components}} 
			\end{algorithmic}
		\end{algorithm}
		\item \textbf{Reading Module (ReadMod)}. \emph{ReadMod} is in charge of reading security policies from the dedicated Block RAM trusted memory and extracting security parameters to be transmitted to the \emph{Checking Module}. For a single 32-bit word, a reading buffer is filled in 1 clock cycle. Then, security parameters are transmitted to the \emph{Checking Module} in 1 additional cycle.
		\item \textbf{Checking Module (CheckMod)}. For a simple implementation, \emph{Checking Module} verifies two parameters: read/write access right and data format thanks to a set of comparators and a test function (see Figure \ref{fig6}).\newpage
		\begin{figure}[htbp]
			\centering
			\includegraphics[width=.75\textwidth]{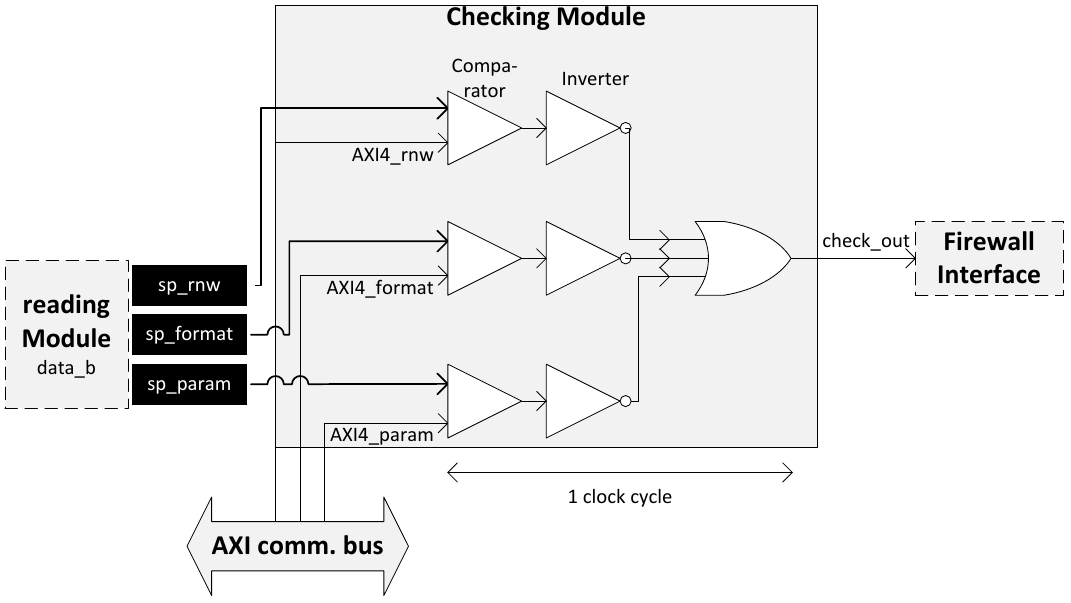}
			\caption{Structure of a Checking Module}
			\label{fig6}
		\end{figure}
		~\\
		First, there is a preliminary test on the value of the \emph{ARID} signal extracted from the bus: when this signal is not equal to 0, the data being analyzed is associated with a read; otherwise (\emph{ARID=0}), it is related with a write operation. This test function output is used to check two parameters:
		\begin{itemize}
			\item The select input of a multiplexer for format value verification (\emph{ARSIZE} and \emph{AWSIZE} contain format values for read and write operations \cite{Arm12}).
			\item Then, access right and data format are compared with \emph{sp\_rnw} and \emph{sp\_format} values extracted from the security policy. As all the parameters are instantiated in parallel (see Figure \ref{fig6}), all the parameters are analyzed at the same time. In Figure \ref{fig6}, inverters and the OR gate produce a global signal representative of all the controls done in the \emph{Checking Module}: if one or more comparators go wrong, the final output will be 0 (in the other case, when all comparators go fine, the output is 1).
		\end{itemize}
		\item \textbf{Finite State Machine (FSM)}: the FSM managing firewall behavior.
	\end{itemize}
	\begin{figure}[htbp]
		\centering
		\includegraphics[width=.324\textwidth]{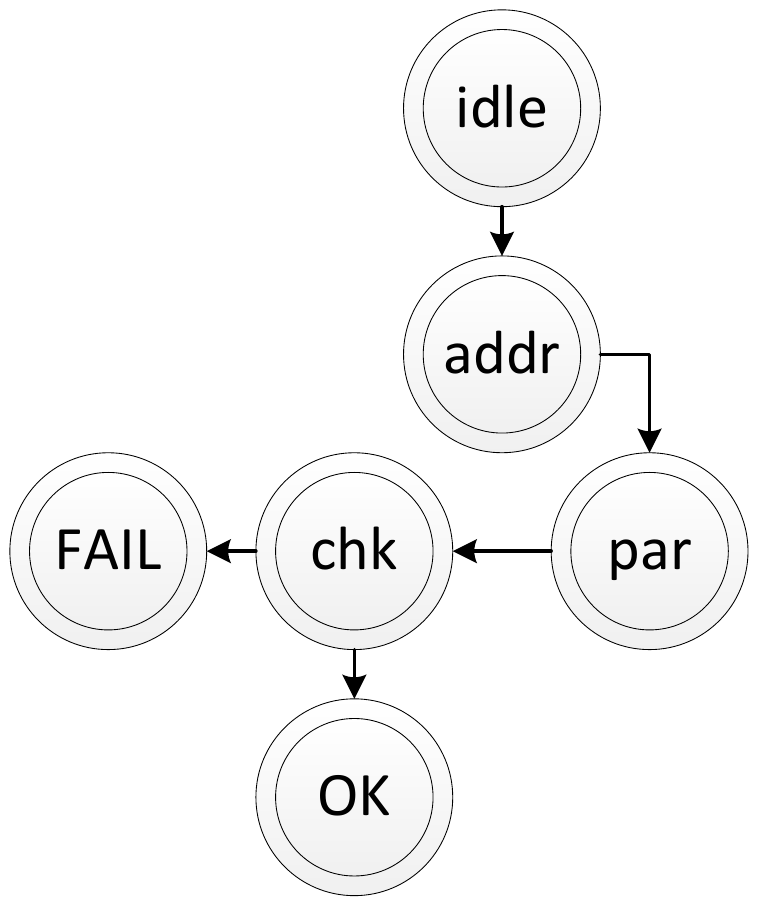}
		\caption{FSM associated with the Security Builder module}
		\label{fig7}
	\end{figure}
	~\\
	FSM described in Figure \ref{fig7} represents the simplified behavior of a \emph{Security Builder} embedded in a \emph{Local Firewall} (see Figure \ref{fig3}). It contains the main states of this module (without error procedure or cryptographic feature):
	\begin{itemize}
		\item \emph{Idle} state. The \emph{Security Builder} module waits for a new incoming data to be analyzed (and received from the \emph{Firewall Interface}).  
		\item \emph{Addr} state. Allowance to recover Block RAM address where data is located.
		\item \emph{Par} state. Security policy reading step
		\item \emph{Chk} state. Checking operation performed in 2 clock cycles: 1 cycle for the preliminary test and another cycle for the combinatorial part (comparators and gates). If parameters match with the security policy, \emph{OK} state is enabled (otherwise, \emph{FAIL} state).
	\end{itemize}
	In order to have a complete firewall behavior, the final step is the transmission of the \emph{check\_out} signal to the \emph{Firewall Interface} sub-module: if checking operations match with the security policy, signals are transmitted in a synchronized manner; otherwise, data are not transmitted (the process resulting from an error is described later in this paper). Checking a 32-bit data word takes 4 clock cycles. Therefore, checking N 32-bit data words is done in 4N clock cycles in a non-pipelined architecture.
	
	\emph{Local Firewalls} allow to monitor and to filter any communication on the bus. They do not monitor the internal execution flow of processors and their associated cache behaviors. If an attacker sets up a cache based attack (e.g. trace-driven or time-driven), \emph{Local Firewalls} will not block such an attack as it does not correspond to illegal operations (i.e. access right, data format). Additional solution (e.g. software-based solution) should be used to provide some countermeasures against this type of attacks. If an attacker tampers an application code that is not protected with integrity mechanisms and if this malicious code tries to perform illegal accesses then \emph{Local Firewalls} will detect and block this attack. \emph{Local Firewalls} do not protect against side channel attacks but allow to detect and stop any attack that does not respect the security policies associated with an application. Extending these mechanisms with cryptography properties enables to address a large class of potential attacks.
	
	\subsection{Cryptographic Firewall}
\begin{figure}[htbp]
		\centering
		\includegraphics[width=.75\textwidth]{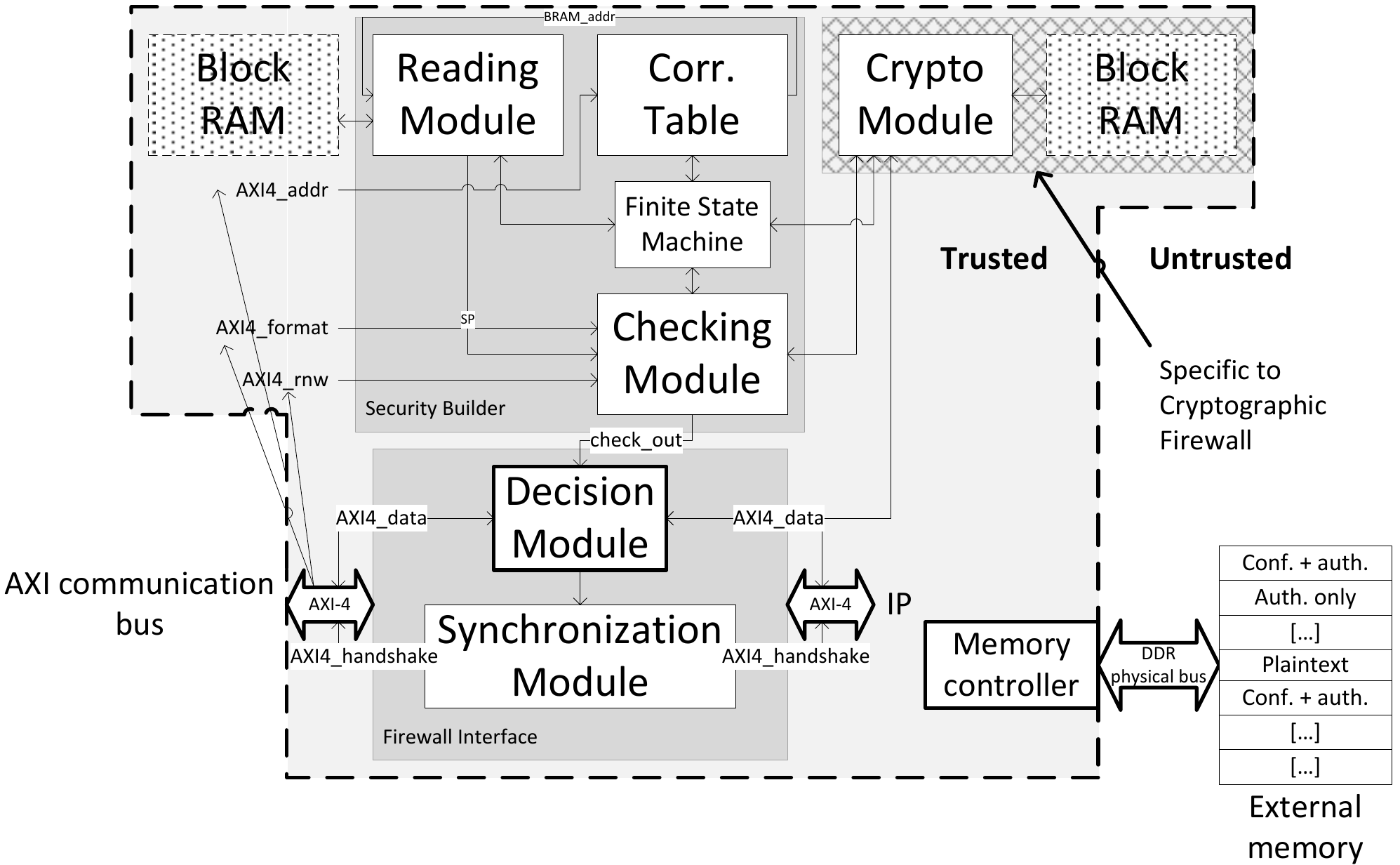}
		\caption{Structure of a Cryptographic Firewall}
		\label{fig8}
	\end{figure}

\begin{figure}[htbp]
		\centering
		\includegraphics[width=.75\textwidth]{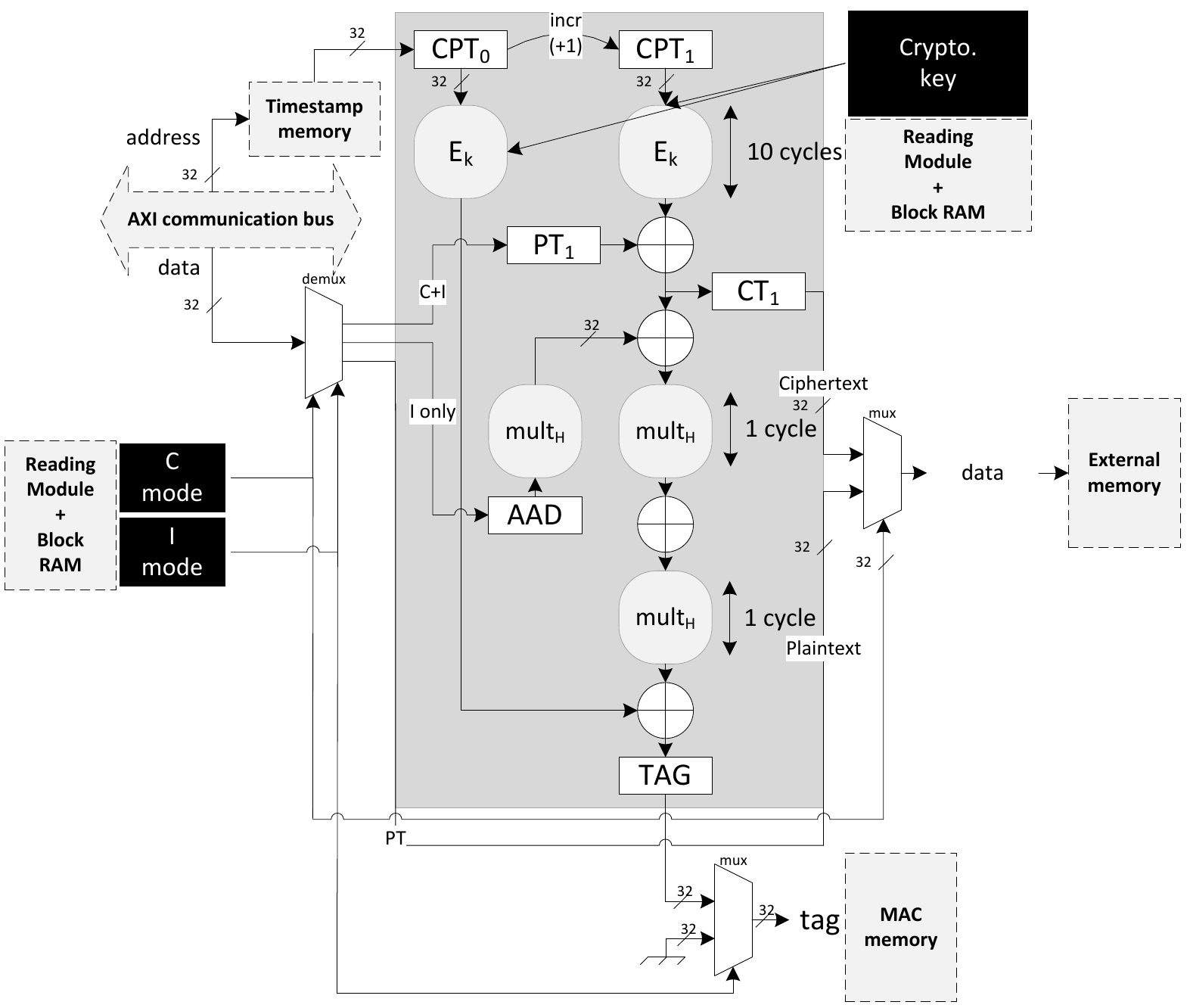}
		\caption{Cryptographic Module: AES-GCM}
		\label{fig9}
	\end{figure}

	The external memory controller has a dedicated firewall offering cryptographic features in addition to security services embedded in a \emph{Local Firewall}. This type of firewall (also known as \emph{Cryptographic Firewall}) is able to protect data stored in the external memory in terms of confidentiality and authentication. The overall structure of a \emph{Cryptographic Firewall} is shown in Figure \ref{fig8}.
	~\\
	The main difference between a \emph{Local} and a \emph{Cryptographic Firewall} is the \emph{Security Builder}. Datapath is modified to take into account confidentiality and integrity features implemented in a \emph{Cryptographic Firewall}:
	\begin{itemize}
		\item In case of a read: the first step is to get the security policy associated with data currently analyzed. Then, data is processed by the cryptographic module (deciphering and eventually authentication checking) before being sent to the \emph{Security Builder} (verification of access rights, data format\ldots).
		\item In case of a write: datapath goes through the \emph{Security Builder} before going to the cryptographic module if parameters match (\emph{check\_out=1}).
	\end{itemize}
	In this contribution, firewalls have to operate with a low-latency feature. Therefore, a global protection of external memories (in terms of confidentiality and authentication) is not the best option. That is the reason why we want to implement an algorithm offering confidentiality and authentication separately (the security mode is specified by the security policies). Four options are considered in this work:
	\begin{itemize}
		\item AES + MD5. The simplest solution uses the AES standard for confidentiality and a hash function such as MD5 for integrity. Even if it has been developed in the early 90s, MD5 is still widely used \cite{Jarvinen05,Shi2012}.
		\item AES + SHA-2. Developed when security weaknesses were discovered in MD5. According to several surveys\footnote{\href{http://eprint.iacr.org/2004/207.pdf}{http://eprint.iacr.org/2004/207.pdf}}, other weaknesses were highlighted and implementations with better security were developed.
		\item AES + SHA-3. NIST\footnote{National Institute of Standards and Technology} launched a contest in 2007 to define SHA-3. The winner was announced in October 2012, it is a protocol developed by STMicroelectronics and NXP named Keccak\footnote{\href{http://keccak.noekeon.org/}{http://keccak.noekeon.org/}}. For experimentations, we use results extracted from the ATHENA database created by George Mason University\footnote{\href{http://cryptography.gmu.edu/athenadb/fpga_hash/table_view}{http://cryptography.gmu.edu/athenadb/fpga\_hash/table\_view}} for implementations on Xilinx Virtex-6 FPGAs.
		\item AES-GCM. This particular mode of AES is able to perform authenticated ciphering. Confidentiality is guaranteed by plaintext ciphering in CTR mode and authentication is performed by a MAC calculation. \cite{Crenne13} proposes a comparison of various AES modes and a detailed analysis of AES-GCM mode.
	\end{itemize}
	In order to choose the best option in our context of low latency, two metrics are used: latency on a 32-bit data block and throughput of a basic implementation. Results associated to each option are shown in Table \ref{tab2}.
	\begin{table}[htbp]
		\centering
		\caption{Modes comparison for the cryptographic block}
		\label{tab2}
		\begin{tabular}{c|c|c|}
			\cline{2-3}
			& {\bf \begin{tabular}[c]{@{}c@{}}Latency\\ (\# of cycles)\end{tabular}} & {\bf Throughput} \\ \hline
			\multicolumn{1}{|c|}{{\bf \begin{tabular}[c]{@{}c@{}}AES + MD5\\ \cite{Jarvinen05,Shi2012}\end{tabular}}} & 90 & up to 725 Mbits/s \\ \hline
			\multicolumn{1}{|c|}{{\bf \begin{tabular}[c]{@{}c@{}}AES + SHA-2\\ \cite{Chaves06}\end{tabular}}} & 74 & up to 1.8 Gbits/s \\ \hline
			\multicolumn{1}{|c|}{{\bf \begin{tabular}[c]{@{}c@{}}AES + SHA-3\\ (Keccak)\end{tabular}}} & 25 & around 30 Gbits/s \\ \hline
			\multicolumn{1}{|c|}{{\bf \begin{tabular}[c]{@{}c@{}}AES-GCM\\ \cite{McGrew05}\end{tabular}}} & 25 & 30 Gbits/s \\ \hline
		\end{tabular}
	\end{table}
	~\\
	According to Table \ref{tab2}, the best option is AES-GCM (four times faster than solutions based on hash functions). The structure of this mode is described in Figure \ref{fig9}. According to these first results, Keccak (SHA-3 standard) has also interesting performances and could be considered as another alternative.\newpage
	According to Figure \ref{fig9}, a single module is able to perform several options (\zg confidentiality and authentication\zd, \zg authentication only\zd or \zg plaintext\zd). Multiplexors and demultiplexors ports are implemented in order to take into account the security policy parameters defining the confidentiality and authentication properties (respectively \emph{Cmode} and \emph{Imode}). It allows the block to take different datapaths producing ciphertext and tag (and even bypass the whole module if no cryptographic feature is enabled). Keys used for encryption/decryption are extracted from the security policies and sent by the \emph{Reading Module} (as all information related to MACs).
	
	In AES-GCM algorithm, confidentiality is performed by plaintext ciphering. AES function (in CTR mode) in \emph{Ek} blocks generates keystreams (with the key extracted from the security policies stored in a trusted on-chip memory). A timestamp value is used as an input of the AES-GCM to avoid replay attacks. Finally, a XOR is performed with the keystream and the plaintext to produce the ciphertext which can be transmitted to the external memory controller. Authentication is made with a MAC based on universal hash. AAD is a data that can potentially be authenticated without being ciphered.
	
	In Figure \ref{fig9}, encryption of a 32-bit data (\emph{Ek} uses a 128-bit key and a 128-bit data containing the data block to be ciphered on 32 bits and padded with zeros) is performed in 10 clock cycles and authentication in 2 additional cycles (2 Galois multipliers MULTH are needed \cite{Crenne13,Crenne11a}). The overall latency for a set of N 32-bit data protected in terms of confidentiality and authentication is given by Equation \ref{eq1}.
	\begin{equation}
	latency(N)=10+(10+2)*N
	\label{eq1}
	\end{equation}
	Finally, using AES-GCM allows firewalls to provide low-latency cryptographic features to the final user while keeping flexibility in terms of available modes (confidentiality and authentication). In particular for authentication, AES-GCM takes 2 clock cycles while MD5 (respectively SHA-2) takes 64 (respectively 80) cycles to do so. Tag produced by the AES-GCM core is not ciphered as it is stored in a trusted Block RAM memory. As Block RAMs are dual-port memories, one port is left free for further parameters update operations (these operations are described in Section \ref{sec_dynamic}).
	
	\section{Update services for dynamic hardware firewall configuration}\label{sec_dynamic}
	\subsection{Security leakages in static security}
	Section \ref{sec_static} proposes a solution to protect a multiprocessor architecture with static security enhancements. Unfortunately, this is a solution where security policies cannot be updated in case of an attack. At this step, firewalls only detect an attack thanks to the \emph{Security Builder}. Therefore, mechanisms are needed to update firewall rules without creating security weaknesses within the system. Security update can be done by partial or complete reconfiguration (download of a partial or complete bitstream). However, the solution proposed to update the security policies in this section aims to have the following features:
	\begin{itemize}
		\item No system blocking. System behavior should not be stopped during security updates. Furthermore, malicious data must not leak during this process.
		\item Low-latency update. Security must be updated as fast as possible.
		\item Different security modes. Firewalls should offer a hierarchy of security levels in order to allow mechanisms to be set in a more or less permissive mode. It is assumed that security levels are defined by access rights (read/write, read-only and so on).
	\end{itemize}
	
	\subsection{Security policies evolution}
	Two categories of components are defined according to their ability to handle confidential information. For instance, critical IPs such as encryption blocks must not reveal information in case of an attack. If an attacker is able to extract cryptographic keys, it would be a major failure for the embedded system. In this case, critical IPs must be placed in a quarantine mode where no read or write are allowed. For non-critical IPs, an intermediate mode is defined where only read accesses are allowed. This feature allows backing up firewalls configuration and confidential information before eventually switching to the quarantine mode.
	
	In the most critical case, when an attack is detected by a firewall already set to the quarantine mode, it is assumed that the system must be completely reset. Therefore, the initial bitstream (containing initial security policies) is downloaded into the FPGA chip to ensure a secure execution environment to the system.\newpage\noindent When a firewall has to be updated, there are two potential areas to be modified:
	\begin{itemize}
		\item \emph{Correspondence Table} module: it defines the relationship between the address spaces and a related security policy. In this work, it is considered that all the address spaces of each IP are covered by, at least, one security policy. Therefore, the only component to be updated is the memory containing the security policies as explained below.
		\item Block RAM memory containing the security policies. These memories contain the rules to be verified for each communication and memory access.
	\end{itemize}
	For both \emph{Local} and \emph{Cryptographic} firewalls, information related to read/write access rights is stored in a 32-bit word in a Block RAM memory. As a result, updating a security policy means writing a 32-bit word (with new access rights values) in a BRAM; this is done in one clock cycle\footnote{\href{http://www.xilinx.com/support/documentation/ip\_documentation/axi\_bram\_ctrl/v1\_03\_a/ds777\_axi\_bram\_ctrl.pdf}{http://www.xilinx.com/support/documentation/ip\_documentation/axi\_bram\_ctrl/v1\_03\_a/ds777\_axi\_bram\_ctrl.pdf}}. Finally, updating N security policies in one firewall is performed according to the following equation:
	\begin{equation}
	duration(N)=N~cycles
	\label{eq2}
	\end{equation}
	\subsection{Update services architecture}
\begin{figure}[htbp]
		\centering
		\includegraphics[width=.71\textwidth]{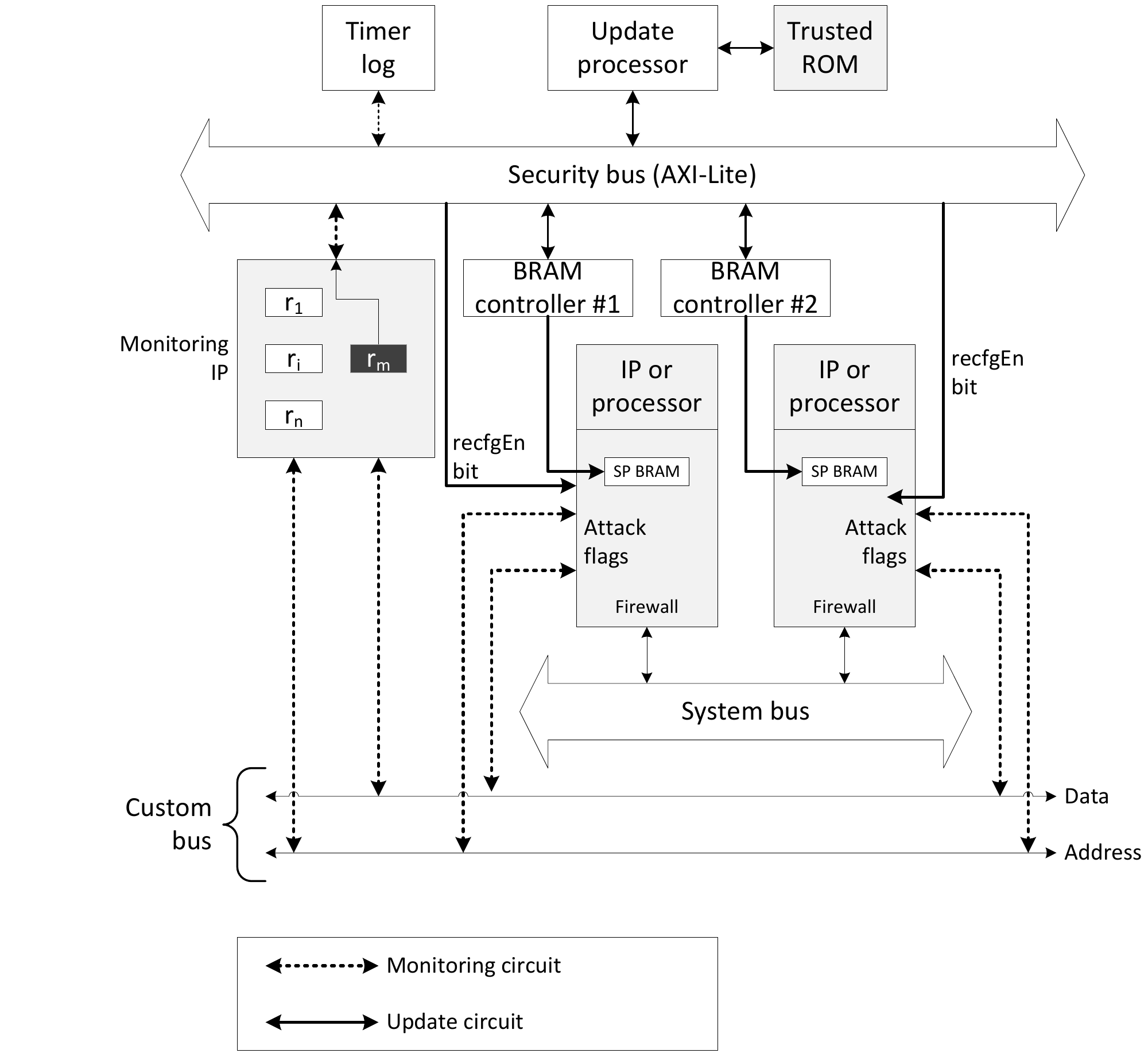}
		\caption{Global architecture of the monitoring area and update services}
		\label{fig10}
	\end{figure}

	When an attack is detected, Block RAMs must be updated with the new security policies in order to keep a secure execution environment. All the components are connected to an AXI-Lite communication bus (labeled \zg security bus\zd{} in Figure \ref{fig10}). A dedicated processor (labeled \zg update processor\zd{} in Figure \ref{fig10}) keeps records of all important events in a log file (timestamps, attacks and update progress).
	~\\
	Each firewall has a connection with a monitoring IP through a dedicated bus. While this IP and a timer are used for attack detection and reporting, BRAM controllers are used to update the security policies embedded in each firewall BRAM memory. As the code of the \zg update processor\zd is stored in a trusted on-chip memory (as the architecture is composed of the security bus, timer log and custom bus), this update process cannot be tampered by malicious accesses.
	
	\subsubsection{Monitoring IP}
	The first operation performed by the architecture presented in Figure \ref{fig10} is the monitoring of attack events thanks to a dedicated IP (labeled \zg Monitoring IP\zd in Figure \ref{fig10}). In order to detect attacks, three flags are extracted from firewalls (both \emph{Local} and \emph{Cryptographic}):
	\begin{itemize}
		\item \emph{checkFlag} (cF): one of the controls (read/write access right or data format) failed in the \emph{Checking Module}. This signal is similar to the \emph{check\_out} signal described in Section \ref{sec_static}.
		\item \emph{notFoundFlag} (nF): attempt to access an unknown address (i.e. an address not included in the \emph{Correspondence Table}).
		\item \emph{authenticationFlag} (aF): authentication verification failed (flag specific to a \emph{Cryptographic Firewall} linked to the external memory controller).
	\end{itemize}
	\begin{figure}[htbp]
		\centering
		\includegraphics[width=.75\textwidth]{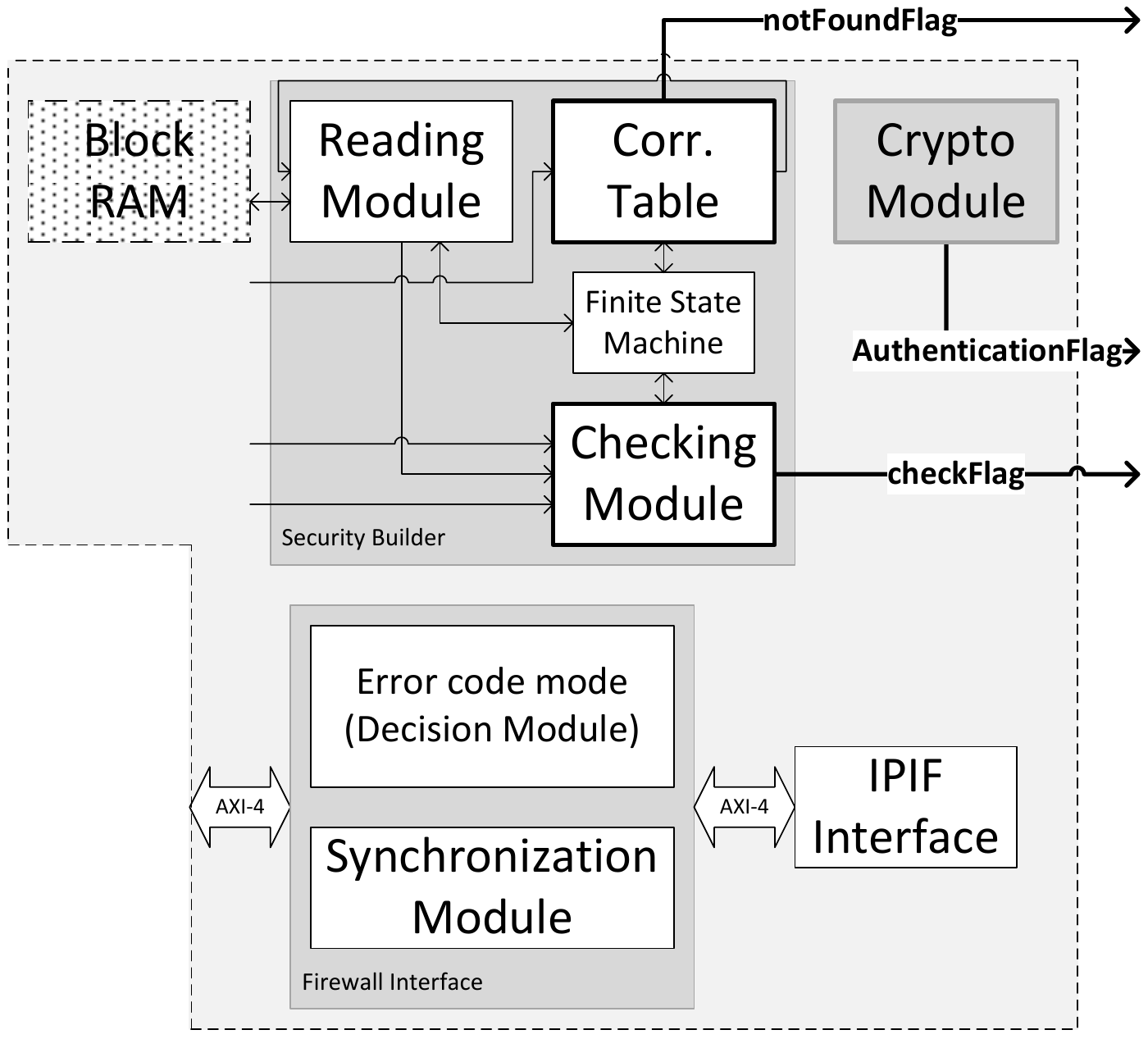}
		\caption{Generic structure of a firewall with flags extraction}
		\label{fig11}
	\end{figure}
	Figure \ref{fig11} shows a mixed structure of a \emph{Local Firewall} and a \emph{Cryptographic Firewall}. Each flag is stored on a significant bit in internal registers of the monitoring IP (each IP has a dedicated register).\newpage
	\begin{figure}[htbp]
		\centering
		\includegraphics[width=.75\textwidth]{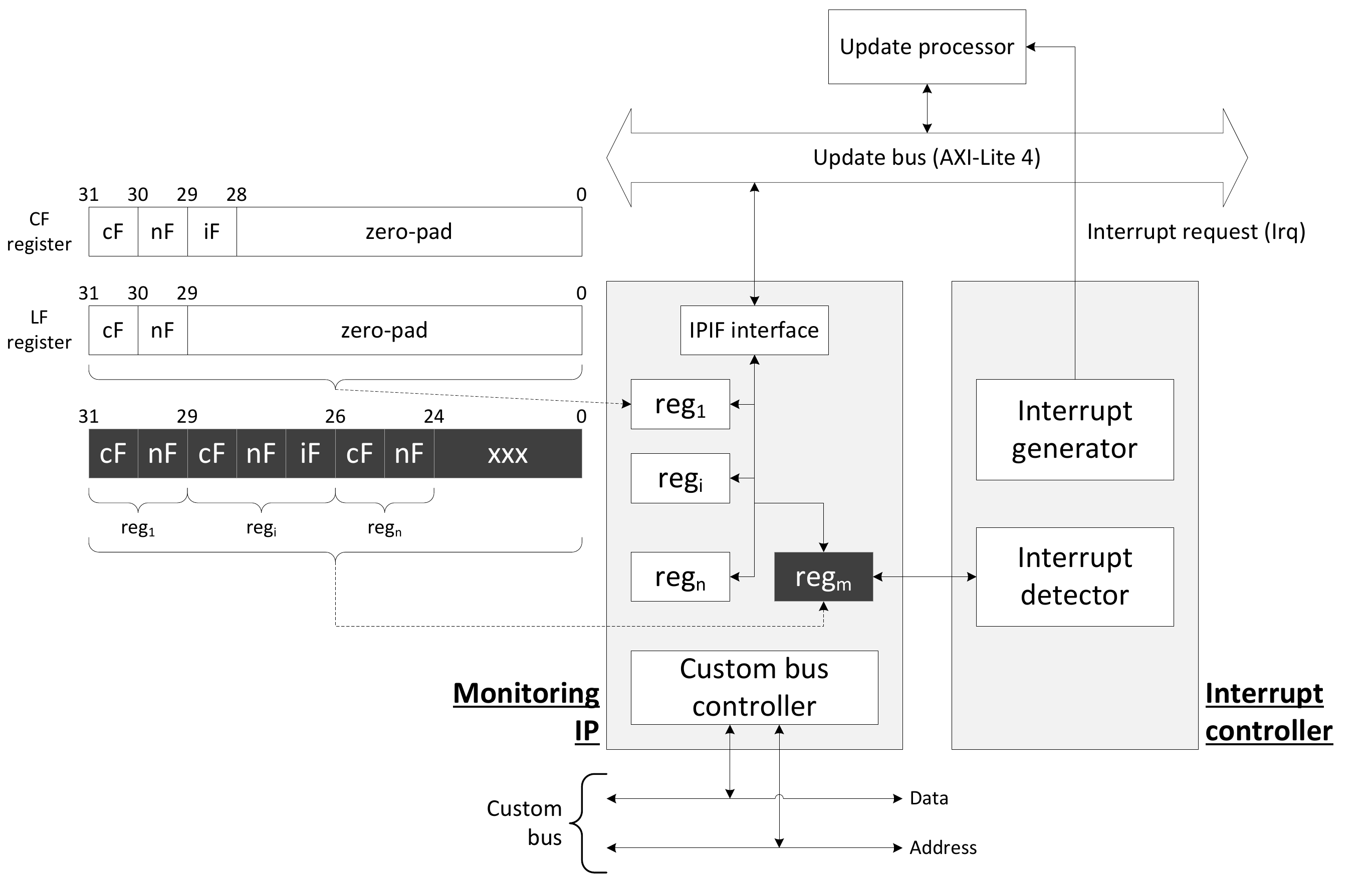}
		\caption{Architecture of the monitoring IP and interrupt routine}
		\label{fig12}
	\end{figure}
	Figure \ref{fig12} shows the detailed structure of the monitoring IP. Each firewall provides its flag information through a custom bus (address and data). Each $reg_i$ is a 32-bit register with only 3 significant bits (corresponding to 3 flags). The main register $reg_m$ concatenates all the significant bits. As $reg_m$ register is also on 32 bits, the monitoring IP can manage up to 10 firewall registers. An interruption controller reads the main register $reg_m$ and sends an interrupt request to the update processor as soon as one of the $reg_m$ bits is equal to zero (i.e. when an error is detected by a firewall). This update processor (executing the interrupt routine) has several features:
	\begin{itemize}
		\item This processor is aware of the security context: security modes (see Section \ref{sec_static}) are known for each firewall.
		\item When a firewall has to be updated, security policies are set to a less permissive configuration.
	\end{itemize}
	Beyond these monitoring tasks, this processor is also in charge of the security policies update (i.e. writing new values in Block RAMs linked to firewalls).
	
	\subsubsection{Security update protocol}
\begin{figure}[htbp]
		\centering
		\includegraphics[width=.75\textwidth]{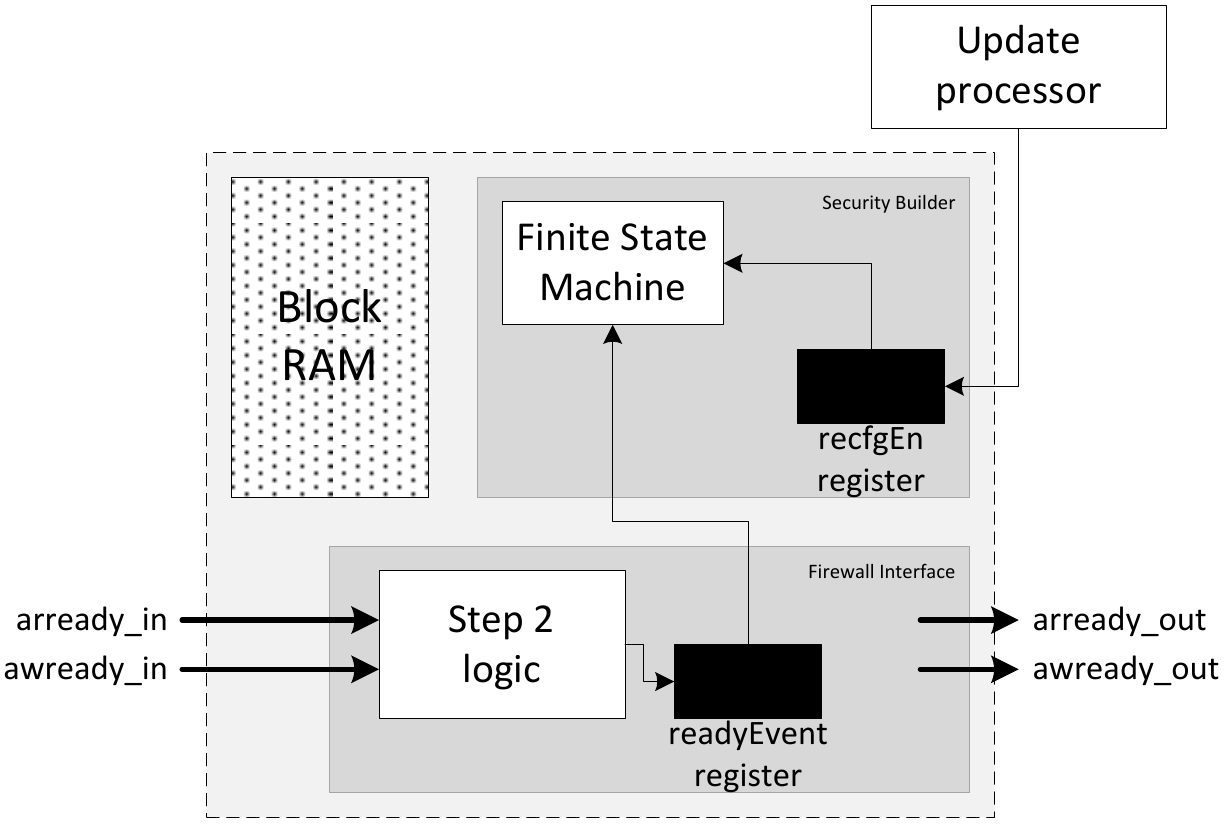}
		\caption{Partial architecture of a firewall with update process information}
		\label{fig13}
	\end{figure}

	Compared to the static protection described in Section \ref{sec_static}, firewall structure is slightly modified. Both Block RAM data ports are used (one directly connected to the firewall, the other connected to the update processor). Therefore, it may happen that both data ports are used at the same time to access the same memory location. The architecture presented in Figure \ref{fig13} has a mechanism, based on communication properties to avoid such cases.
	
	MPSoCs considered in this work use the AXI protocol from ARM as the communication bus. Transactions in AXI protocol are based on a \zg handshake\zd{} protocol with couples of $valid/ready$ signals for address and data channels (for both read and write \cite{Arm12}). Output information is only available when both signals (\emph{valid} and \emph{ready}) are in their high state (no information is transmitted in other cases). All the transactions using the AXI communication protocol are performed in two steps: a handshake on addresses is performed before a handshake on data signals.
	
	Therefore, a mechanism is implemented in each \emph{Firewall Interface} module to avoid concurrent accesses to BRAM memory location by both firewall and update processor. When an attack is detected in a firewall, output ready signals of the \emph{Firewall Interface} are kept in their low state. This step aims to \zg freeze\zd{} the communication bus and prevent malicious accesses during the update process. Then, the update process is executed in accordance with the simplified architecture shown in Figure \ref{fig13}.
	~\\
	First, security update is enabled by the update processor writing a \zg 1\zd{} value in the \emph{recfgEn} register embedded in the \emph{Security Builder}. Thus, the FSM associated to this process goes from a monitoring state to an update state. Then, while the processor retrieves the new security policy to be written into the memories, all transactions are blocked as previously described using the \zg handshake\zd{} feature. If a rising edge occurs on a ready signal ($arready\_in$ or $awready\_in$), it is blocked until the update process is finished (i.e. $arready\_out=0$ and $awready\_out=0$ while $recfgEn=1$) and a \zg 1\zd value is written into the \emph{readyEvent} register embedded in the \emph{Firewall Interface} module. When the new security policy has been written, the update processor writes a \zg 0\zd value in the \emph{recfgEn} register to notify the end of the process.
	
	Finally, according to the \emph{readyEvent} register value, data blocked in the input port of the Firewall Interface is analyzed with the new security policy only if $readyEvent=1$ (i.e. if a rising edge on a ready signal has been detected during update). The overall latency of the update process depends on the number of security policies to be modified. This process usually contains five steps:
	\begin{itemize}
		\item Extracting flags from firewalls to the monitoring IP and blocking of firewalls. This step is done in 1 clock cycle.
		\item Interruption routine execution. An interrupt request is sent in 2 cycles.
		\item Computation of the new security configuration (performed in software). It corresponds to the software latency on the update processor. For a basic implementation, the new security policy is retrieved in 148 clock cycles. Nevertheless, other algorithms may be considered. In this case, response time would be impacted.
		\item Writing the new security policy. This step depends of the number of security policies to be written. Results are given in Section \ref{sec_res}.
		\item Reactivation of the main application after update completeness. This is done in 1 clock cycle.
	\end{itemize}
	
	\section{Implementation results}\label{sec_res}
	\subsection{Experimental setup}\label{subsec_set}
\begin{figure}[htbp]
		\centering
		\includegraphics[width=.75\textwidth]{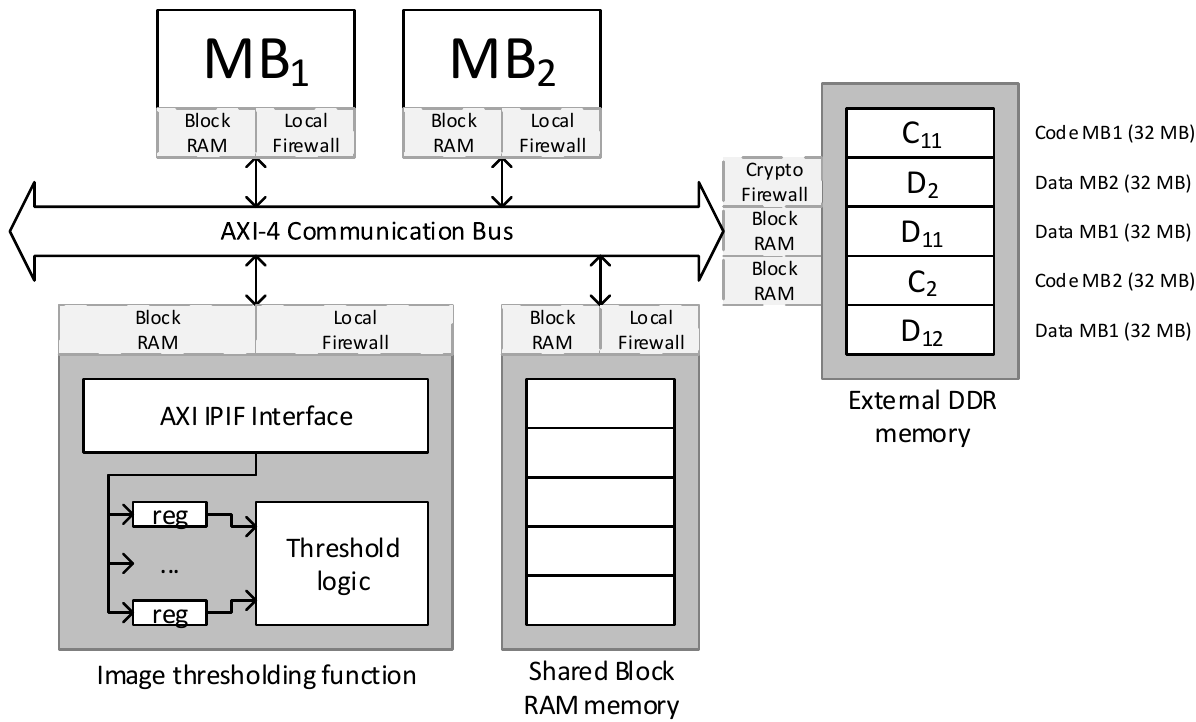}
		\caption{Case study architecture}
		\label{fig14}
	\end{figure}

\begin{figure}[htbp]
		\centering
		\includegraphics[width=.75\textwidth]{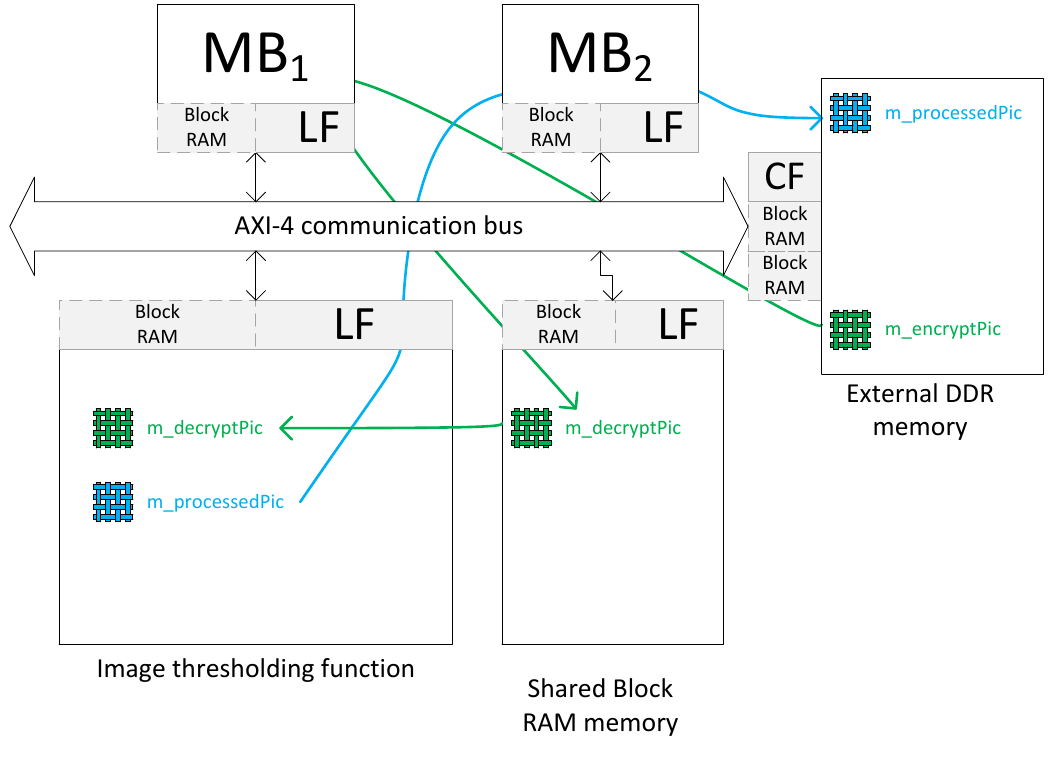}
		\caption{Illustration of picProc on the case study}
		\label{fig15}
	\end{figure}

	In order to validate this contribution, implementations and simulations are done on a case study representative of a real embedded system on a Xilinx ML605 development board (including a Virtex-6 FPGA). We have developed a system performing image processing operations. Its architecture is shown in Figure \ref{fig14}, it contains 2 MicroBlaze softcore processors and 2 IPs:
	\begin{itemize}
		\item An image processing IP. This IP contains several programmable registers and performs a threshold function on a picture.
		\item It is assumed that temporary pictures and other information (code, user profiles and so on) are stored in a shared Block RAM memory.
	\end{itemize}
	~\\
	It is also assumed that the external memory is shared between the 2 processors and split into memory sections of 32 MB each:
	\begin{itemize}
		\item Processor MB1: 1 code section (C11) and 2 data sections (D11 and D12).
		\item Processor MB2: 1 code section (C21) and 1 data section (D21).
	\end{itemize}
	Regarding cryptographic options, memory sections can be protected in \zg confidentiality and integrity\zd{} (C11 and D11), \zg integrity only\zd{} (D12) or even in plaintext (C21 and D21). All transactions are based on a 32-bit data format. Finally access rights are defined in Table \ref{tab3}.\newpage
	\begin{table}[htbp]
		\centering
		\caption{Access rights for the case study}
		\label{tab3}
		\begin{tabular}{c|c|c|c|c|}
			\cline{2-5}
			& \multirow{2}{*}{{\bf \begin{tabular}[c]{@{}c@{}}Shared\\ memory\end{tabular}}} & \multirow{2}{*}{{\bf \begin{tabular}[c]{@{}c@{}}Image\\ processing\end{tabular}}} & \multicolumn{2}{c|}{{\bf External memory}} \\ \cline{4-5} 
			&  &  & C21, D21 & C11, D11, D12 \\ \hline
			\multicolumn{1}{|c|}{{\bf MB1}} & Read only & Read/write & No access & Read/write \\ \hline
			\multicolumn{1}{|c|}{{\bf MB2}} & Read/write & Write only & Read/write & No access \\ \hline
		\end{tabular}
	\end{table}
	These security policies are built to illustrate all the cases (in terms of policies and cryptographic options) on a multiprocessor architecture implemented on a FPGA chip. Then, the case study has some requirements on the software layer (for both applications and operating
	systems):
	\begin{itemize}
		\item On the one hand, there are some requirements on the operating system. The \emph{standalone} OS provided by Xilinx development tools is used: it does not take into account multitask aspects but allows making measurements on communications between elements of the system. Consequently, a POSIX\footnote{\href{http://standards.ieee.org/develop/wg/POSIX.html}{http://standards.ieee.org/develop/wg/POSIX.html}} compliant operating system is used to run custom applications \emph{picProc}, \emph{picDRM} and \emph{picDec}. Xilkernel (from Xilinx) is chosen but $mu$COS\footnote{\href{http://micrium.com/rtos/ucosii/overview/}{http://micrium.com/rtos/ucosii/overview/}} or $mu$CLinux\footnote{\href{http://www.uclinux.org/}{http://www.uclinux.org/}} could have been considered as well.
		\item On the other hand, custom applications are developed to illustrate all the situations (access to memories, communications between processors\ldots).
	\end{itemize}
	The main application \emph{picProc} is based on a sample JPEG image. All operations are saved in a log file. \emph{picProc} is executed through the following steps:
	\begin{itemize}
		\item First, $MB_1$ reads the image from the external memory ($m\_encryptPic$) and processes a software deciphering.
		\item Then, the plaintext image $m\_decryptPic$ is stored in the shared BRAM memory.
		\item The next step is the transfer of the plaintext image to the threshold IP. This is performed again by processor $MB_1$.
		\item Finally, processor $MB_2$ writes the threshold image ($m\_processedPic$) in plaintext into the external memory.
	\end{itemize}
	The picProc application process is summarized in Figure \ref{fig15}.
	~\\
	Beyond that, other applications are used to evaluate the case study:
	\begin{itemize}
		\item \emph{picDRM} is an application ran by a processor checking DRM rights of the target image while the other processor does read/write operations.
		\item \emph{picDec} is an application performing a software deciphering of an image by a unique processor. It serves as a reference for latency results as it does not imply a \emph{Cryptographic Firewall}.
	\end{itemize}
	\newpage
	\subsection{Security analysis}\label{subsec_secana}
	\subsubsection{Context}
\begin{figure}[htbp]
		\centering
		\subfloat[This work: distributed with cryptography-enhanced memory interface]{\label{fig16a}
			\includegraphics[width=.31\textwidth]{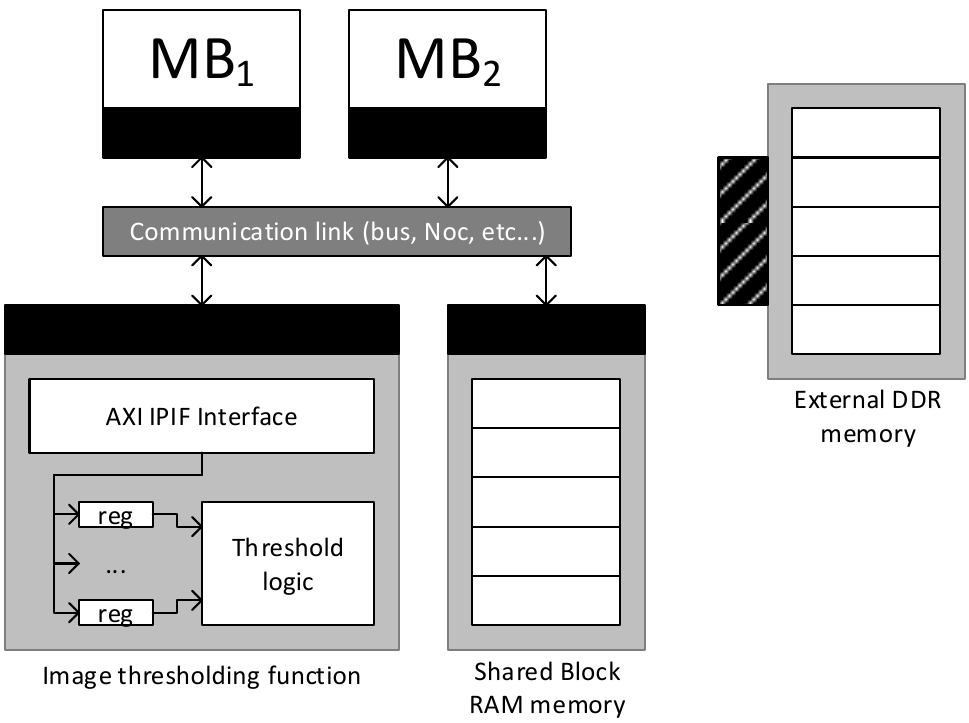}
		}
		\subfloat[\cite{Coburn05} approach: security checking centralized in a specific module]{\label{fig16b}
			\includegraphics[width=.31\textwidth]{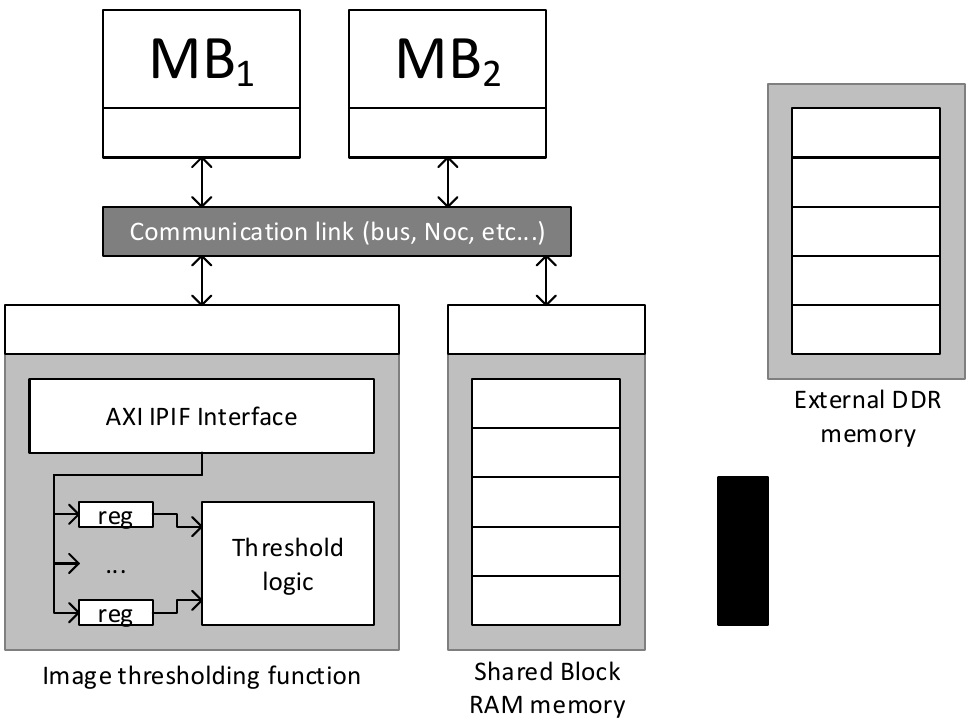}
		}
		\subfloat[\cite{Fiorin08b} approach: distributed with a generic memory interface]{\label{fig16c}
			\includegraphics[width=.31\textwidth]{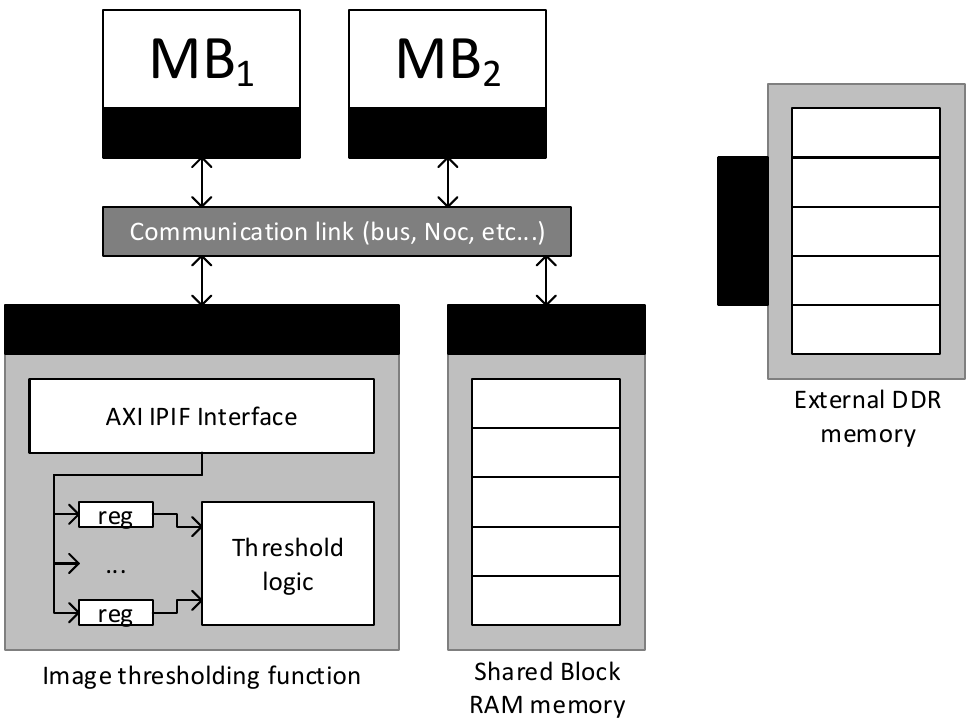}
		}
		\caption{Distributed and centralized behaviors}
		\label{fig16}
	\end{figure}

	The main goal of this section is to provide a security analysis for this work compared to other approaches \cite{Fiorin08b,Coburn05}. This analysis is based on the case study described in the previous section where two attack scenarios will be discussed to verify how implementations behave against the threat model defined in Section \ref{sec_context}. Implementations of \cite{Fiorin08b} and \cite{Coburn05} on the previously defined case study is described in Figure \ref{fig16}.
	~\\
	In Figure \ref{fig16}, it is assumed that the communication network is the one used in each reference. This is due to the fact that in \cite{Fiorin08b,Coburn05} security mechanisms rely partially on the communication architecture. In the original case study, there are memory pages with several cryptographic features.\newpage\noindent This security analysis explores three options for our approach:
	\begin{itemize}
		\item Without cryptographic features (to be fair with existing works that do not provide such services). Therefore, all the data is in plaintext. It would be similar to the case where only \emph{Local Firewalls} are implemented.
		\item With integrity only.
		\item With both confidentiality and integrity (the full AES-GCM core is enabled).
	\end{itemize}
	Then, the two scenarios considered in this analysis are as follows:
	\begin{itemize}
		\item Scenario S1. In a OCR (\emph{Optical Character Recognition}) context, an attacker changes a plaintext data packet with another. For example, a packet containing an image and the address of the next packet is replaced with a noised version and an unknown address. In this case, the threshold function computes an unexpected result.
		\item Scenario S2. Actually, the threshold value is hardcoded in the IP. It is assumed that the attacker was able to either build another IP with another threshold value or reconfigure the system with his malicious IP. As the threshold is different, the result will lead to different results: especially in a OCR context, this attack may dramatically change the application performance.
	\end{itemize}
	Following subsections described how implementations of existing works behave according to these scenarios.
	
	\subsubsection{Scenario S1}
	\emph{With integrity only}. In this case, the modified data is detected by the \emph{Cryptographic Firewall}.\\
	
	\emph{With confidentiality and integrity}. Beyond the fact that data is ciphered, the attack is detected by the \emph{Cryptographic Firewall} with integrity checking.\\
	
	\emph{Without cryptographic services}. For this work, it is considered that a simple \emph{Local Firewall} is attached to the external memory controller for our approach. For both \cite{Fiorin08b} and this work, the malicious packet will be correctly read by processor MB1. Packet will be blocked only at the communication interface of MB1. For \cite{Coburn05}, even if the process is similar, the attack is detected later at the security manager layer which is decentralized in a specific IP. For the second stage of this attack (transmission to a corrupted address), \cite{Fiorin08b} and this work still detect it at MB1 bus interface while \cite{Coburn05} detects it even later within its security manager.
	
	\subsubsection{Scenario S2}
	This scenario is different as a malicious IP is implemented in the system. Therefore, cryptographic features that can be applied to this work (AES-GCM core in the \emph{Cryptographic Firewall}) are not relevant. \cite{Fiorin08b} and \cite{Coburn05} behaves as the previous attack. This work is more relevant as the malicious IP does not provides a \emph{Local Firewall}, monitoring process will be stopped earlier (as soon as data comes in the threshold IP interface): in fact, the update processor (described in Section \ref{sec_dynamic}) is also responsible for monitoring firewalls states. As a consequence, the update processor is aware that the threshold function is performed by an untrusted entity.
	
	\subsubsection{XBox 360 use case}
\begin{figure}[htbp]
		\centering
		\includegraphics[width=.4\textwidth]{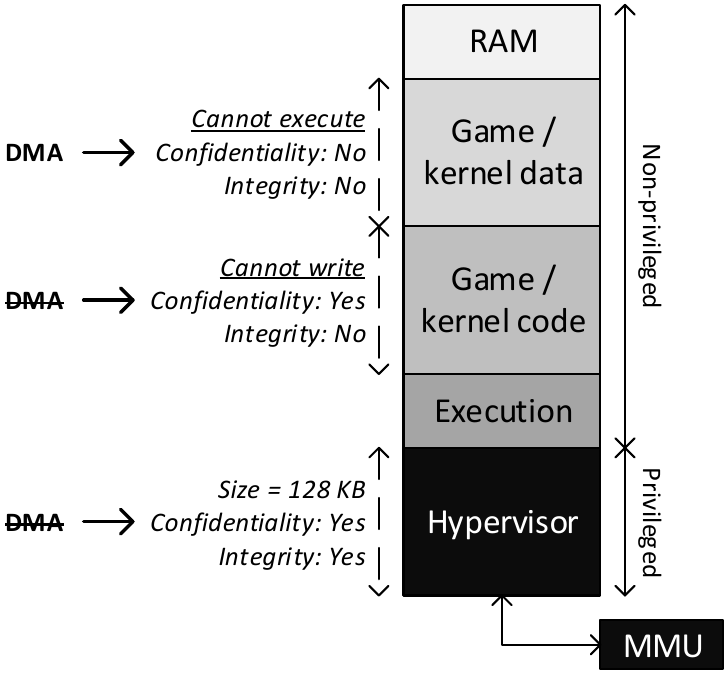}
		\caption{Xbox 360 memory layout}
		\label{fig17}
	\end{figure}

	Security in game consoles has been a hot topic since the 90s with the introduction of modchips for the first generation of Playstation \cite{DeBusschere12}. The XBox 360 is a game console developed by Microsoft in 2005. It was one of the first consoles to take into account existing vulnerabilities aiming to take the control of the platform. XBox 360 was designed to include several security functions. The default memory layout of such a console is presented in Figure \ref{fig17}.
	~\\
	RAM memory is divided in four sections: game/kernel data, game/kernel code, execution section (reserved for reads and writes during runtime) and hypervisor code. Furthermore, the game/kernel data section (which is in plaintext) must be accessed through DMA for specific purposes (basically GPU-related operations) while other memory sections are secured (at least, in terms of confidentiality). In this default configuration, data section is a clear security breaches which can lead to critical exploits such as the \zg King Kong\zd{} attack: this is a well-known method where the attacker uses an event where the GPU access data section in order to get hypervisor-like privileges. In case the system is compromised, the system could accessed unexpected sections. If \emph{Local Firewalls} are implemented in such a system, even if the memory is compromised by one of its memory sections, the approach described in this work could detect the attack in two times:
	\begin{itemize}
		\item As security rules are written in secured on-chip memories, it is considered that accesses to privileged sections can be blocked.
		\item Even if an hypervisor-like user perfoms unwanted operations on other components of the console architecture, abnormal behaviors are detected through \emph{Local Firewalls}.
	\end{itemize}
	Given the fact that console developers are able to implement their architectures on a FPGA chip, the solution described in this work is an efficient solution to block malicious exploits at both hardware and software levels.
	
	\subsection{Experimental results for static and dynamic security}
	\subsubsection{Area results}
	This study is about firewall areas compared to a reference (a MicroBlaze softcore processor running at 100MHz without caches). Firewalls structure was described in Section \ref{sec_static}. The \emph{Correspondence Table} is the only module with a relation between area and security (3 registers are needed to define each security policy).\newpage\noindent It is assumed that the following results consider the worst case (i.e. the largest \emph{Correspondence Table} with 10 security policies). Table \ref{tab4} shows area results of both firewalls (\emph{Local} and \emph{Cryptographic}) compared to the reference MicroBlaze processor.
	\begin{table}[htbp]
		\centering
		\caption{Area results of firewall implementation}
		\label{tab4}
		\begin{tabular}{cc|c|c|c|c|}
			\cline{3-6}
			&  & {\bf Slices} & {\bf Slice regs} & {\bf LUTs} & {\bf\# of Block RAMs} \\ \hline
			\multicolumn{1}{|c|}{\multirow{2}{*}{{\bf \begin{tabular}[c]{@{}c@{}}Local\\ Firewall\end{tabular}}}} & {\bf \begin{tabular}[c]{@{}c@{}}Firewall Interface\\ Security Builder\end{tabular}} & \begin{tabular}[c]{@{}c@{}}76\\ 23\end{tabular} & \begin{tabular}[c]{@{}c@{}}120\\ 3\end{tabular} & \begin{tabular}[c]{@{}c@{}}68\\ 55\end{tabular} & \begin{tabular}[c]{@{}c@{}}0\\ 1\end{tabular} \\ \cline{2-6} 
			\multicolumn{1}{|c|}{} & {\bf Total} & 99 & 123 & 293 & 1 \\ \hline
			\multicolumn{1}{|c|}{\multirow{3}{*}{{\bf \begin{tabular}[c]{@{}c@{}}Crypto\\ Firewall\end{tabular}}}} & {\bf Firewall Interface} & \begin{tabular}[c]{@{}c@{}}76\\ 23\end{tabular} & \begin{tabular}[c]{@{}c@{}}120\\ 3\end{tabular} & \begin{tabular}[c]{@{}c@{}}153\\ 55\end{tabular} & \begin{tabular}[c]{@{}c@{}}0\\ 1\end{tabular} \\ \cline{2-6} 
			\multicolumn{1}{|c|}{} & {\bf Crypto Module} & \begin{tabular}[c]{@{}c@{}}1,166\\ 89.42\%\end{tabular} & \begin{tabular}[c]{@{}c@{}}2,038\\ 94.31\%\end{tabular} & \begin{tabular}[c]{@{}c@{}}2,396\\ 89.10\%\end{tabular} & \begin{tabular}[c]{@{}c@{}}14\\ 93.33\%\end{tabular} \\ \cline{2-6} 
			\multicolumn{1}{|c|}{} & {\bf Total} & 1,304 & 2,161 & 2,689 & 15 \\ \hline
			\multicolumn{2}{|c|}{{\bf MicroBlaze}} & 1,179 & 1,298 & 1,829 & 10 \\ \hline
		\end{tabular}
	\end{table}
	~\\
	One \emph{Local Firewall} has an acceptable area cost compared to one \emph{Cryptographic Firewall} (around 9\%) and to one Microblaze (in this case, nearly 11\%). This is essentially due to the AES-GCM ciphering core: its area corresponds to 90\% of the overall area of one \emph{Cryptographic Firewall}. Therefore, we can also estimate the area of a complete system embedding $x$ \emph{Local Firewalls} and $y$ \emph{Cryptographic Firewalls}. It is assumed that synthesis tools are linear: the area of a structure with 2 identical modules is twice the area of a single module. Areas in terms of slices, registers and LUTs are given in following equations:
	\begin{eqnarray}
	numSlices = 138 \times x + 1,304 \times y\\
	numRegs = 123 \times x + 2,161 \times y\\
	numLuts = 293 \times x + 2,689 \times y
	\end{eqnarray}
	Designer can use these equations to estimate the cost of the hardware firewalls for his/her system. Table \ref{tab5} shows area overheads due to the update mechanisms presented in Section \ref{sec_dynamic}.
	\begin{table}[htbp]
		\centering
		\caption{Synthesis results for the update mechanisms on a Local Firewall}
		\label{tab5}
		\begin{tabular}{cc|c|c|c|c|}
			\cline{3-6}
			\multicolumn{2}{c|}{} & {\bf Slices} & {\bf Slice regs} & {\bf LUTs} & {\bf \# of BRAMs} \\ \hline
			\multicolumn{2}{|c|}{{\bf \begin{tabular}[c]{@{}c@{}}Static solution\\ (Local Firewall)\end{tabular}}} & 138 & 123 & 293 & 1 \\ \hline
			\multicolumn{1}{|c|}{\multirow{4}{*}{{\bf \begin{tabular}[c]{@{}c@{}}Improvements\\ for update\\ solutions\end{tabular}}}} & {\bf Real time} & 6 & 0 & 5 & 0 \\ \cline{2-6} 
			\multicolumn{1}{|c|}{} & {\bf Quarantine} & 17 & 0 & 18 & 0 \\ \cline{2-6} 
			\multicolumn{1}{|c|}{} & {\bf Total overhead} & 5 & 13 & 15 & 0 \\ \cline{2-6} 
			\multicolumn{1}{|c|}{} & {\bf Total overhead} & +20.29\% & +10.57\% & +12.97\% & +0.00\% \\ \hline
		\end{tabular}
	\end{table}
	~\\
	Table \ref{tab5} only shows modifications on a firewall (a complete solution would include a processor, a communication bus\ldots). The \zg improvements for update\zd contribution corresponds to flip-flops used to block the data during the update process (see Section \ref{sec_dynamic}). Otherwise, extensions on a \emph{Local Firewall} leads to a 20\% area overhead (it would be relatively lower on a  \emph{Cryptographic Firewall}). Table \ref{tab6} shows area overheads described in Section \ref{subsec_set}.
	\begin{table}[htbp]
		\centering
		\caption{Area results of several configurations}
		\label{tab6}
		\begin{tabular}{c|c|c|c|c|}
			\cline{2-5}
			& {\bf Slices} & {\bf Regs} & {\bf LUTs} & {\bf \# of BRAMs} \\ \hline
			\multicolumn{1}{|c|}{{\bf \begin{tabular}[c]{@{}c@{}}Solution x0\\ without firewalls\end{tabular}}} & 5,446 & 7,195 & 8,354 & 32 \\ \hline
			\multicolumn{1}{|c|}{{\bf \begin{tabular}[c]{@{}c@{}}Solution x1a with firewalls\\ and without updates\end{tabular}}} & \begin{tabular}[c]{@{}c@{}}7,302\\ +34.08\%\end{tabular} & \begin{tabular}[c]{@{}c@{}}9,848\\ +36.87\%\end{tabular} & \begin{tabular}[c]{@{}c@{}}12,215\\ +46.22\%\end{tabular} & \begin{tabular}[c]{@{}c@{}}51\\ +37.25\%\end{tabular} \\ \hline
			\multicolumn{1}{|c|}{{\bf \begin{tabular}[c]{@{}c@{}}Solution x1b with firewalls\\ and update\end{tabular}}} & \begin{tabular}[c]{@{}c@{}}7,442\\ +1.92\%\end{tabular} & \begin{tabular}[c]{@{}c@{}}9,913\\ +0.66\%\end{tabular} & \begin{tabular}[c]{@{}c@{}}12,405\\ +1.55\%\end{tabular} & \begin{tabular}[c]{@{}c@{}}51\\ +0.00\%\end{tabular} \\ \hline
			\multicolumn{1}{|c|}{{\bf \begin{tabular}[c]{@{}c@{}}Solution x2a with firewalls\\ and without update\end{tabular}}} & +23.38\% & +22.71\% & +32\% & +44.25\% \\ \hline
		\end{tabular}
	\end{table}\newpage\noindent
	Four configurations are taken into account:
	\begin{itemize}
		\item $x0$: this is the original case study without firewalls as shown in Figure \ref{fig14}.
		\item $x1a$: $x0$ architecture with static firewalls as described in Section \ref{sec_static}.
		\item $x1b$: $x1a$ architecture with update improvements (Section \ref{sec_dynamic}).
		\item $x2$a: twice $x1a$ architecture with static firewalls as described in Section \ref{sec_static}: 4 processors, 4 IPs but still one external memory. This configuration is used to illustrate the scalability of the firewalls approach described is this work.
	\end{itemize}
	Case study with static firewalls implies a significant area overhead (34\% in terms of slices): this is mainly due to the AES-GCM core embedded in the \emph{Cryptographic Firewall}. Logic added for update purposes implies an overhead less than 2\% compared to the static version. Regarding the scalability of this approach, the area overhead is up to 24\% in terms of slices for a case study twice larger (configuration $x2a$). It must be noted that the area overhead would proportionally decrease as the architecture grows up (in terms of percentages). Finally, it is possible to make a quantitative and qualitative comparison with \cite{Fiorin08a,Coburn05}. Table \ref{tab7} presents some results.
	\begin{table}[htbp]
		\centering
		\caption{Comparison with existing efforts}
		\label{tab7}
		\begin{tabular}{cc|c|c|c|c|c|}
			\cline{3-7}
			&  & {\bf \cite{Coburn05}} & {\bf \cite{Fiorin08b}} & {\bf \cite{Xilinx12a}} & {\bf \cite{Crenne13}} & {\bf This work} \\ \hline
			\multicolumn{1}{|c|}{\multirow{3}{*}{{\bf \begin{tabular}[c]{@{}c@{}}Standalone\\ monitoring\\ block\end{tabular}}}} & \multirow{2}{*}{{\bf \begin{tabular}[c]{@{}c@{}}Processor\\ ratio (\%)\end{tabular}}} & SEI & DPU & Full & SMM + Ctrl & LF \\ \cline{3-7} 
			\multicolumn{1}{|c|}{} &  & 6.20\% & 25\% & 100\% & 73\% & 11.30\% \\ \cline{2-7} 
			\multicolumn{1}{|c|}{} & {\bf \begin{tabular}[c]{@{}c@{}}Reference\\ (LUTs)\end{tabular}} & N/A & N/A & 1829 & 947 & 293 \\ \hline
			\multicolumn{1}{|c|}{\multirow{4}{*}{{\bf \begin{tabular}[c]{@{}c@{}}Case study\\ overhead\end{tabular}}}} & \multirow{2}{*}{{\bf Reference}} & \multirow{2}{*}{N/A} & \multirow{2}{*}{N/A} & Nude & +1 SMM/Ctrl & +4 LF \\ \cline{5-7} 
			\multicolumn{1}{|c|}{} &  &  &  & +0\% & +11.34\% & +14.03\% \\ \cline{2-7} 
			\multicolumn{1}{|c|}{} & \multirow{2}{*}{{\bf Custom}} & +1 LF & +4 LF & Nude & +1 LF & +4 LF \\ \cline{3-7} 
			\multicolumn{1}{|c|}{} &  & +11.30\% & +14.03\% & +0\% & +11.30\% & +14.03\% \\ \hline
			\multicolumn{2}{|c|}{{\bf \begin{tabular}[c]{@{}c@{}}Update\\ features\end{tabular}}} & No & Yes & No & No & Yes \\ \hline
		\end{tabular}
	\end{table}
	~\\
	In terms of area, this work is less efficient than the SECA solution \cite{Coburn05}: we propose an approach where controls are performed in each firewall while SECA has a centralized security checker. Otherwise, our approach is better than \cite{Fiorin08a} solution: for security update, no additional memory is needed as a communication property is used to block data (\zg handshake\zd{} feature of AXI protocol) while Fiorin et al. approach stores all incoming data in a buffer until update completeness.
	
	\subsubsection{Latency results}
\begin{figure}[htbp]
		\centering
	\includegraphics[width=.75\textwidth]{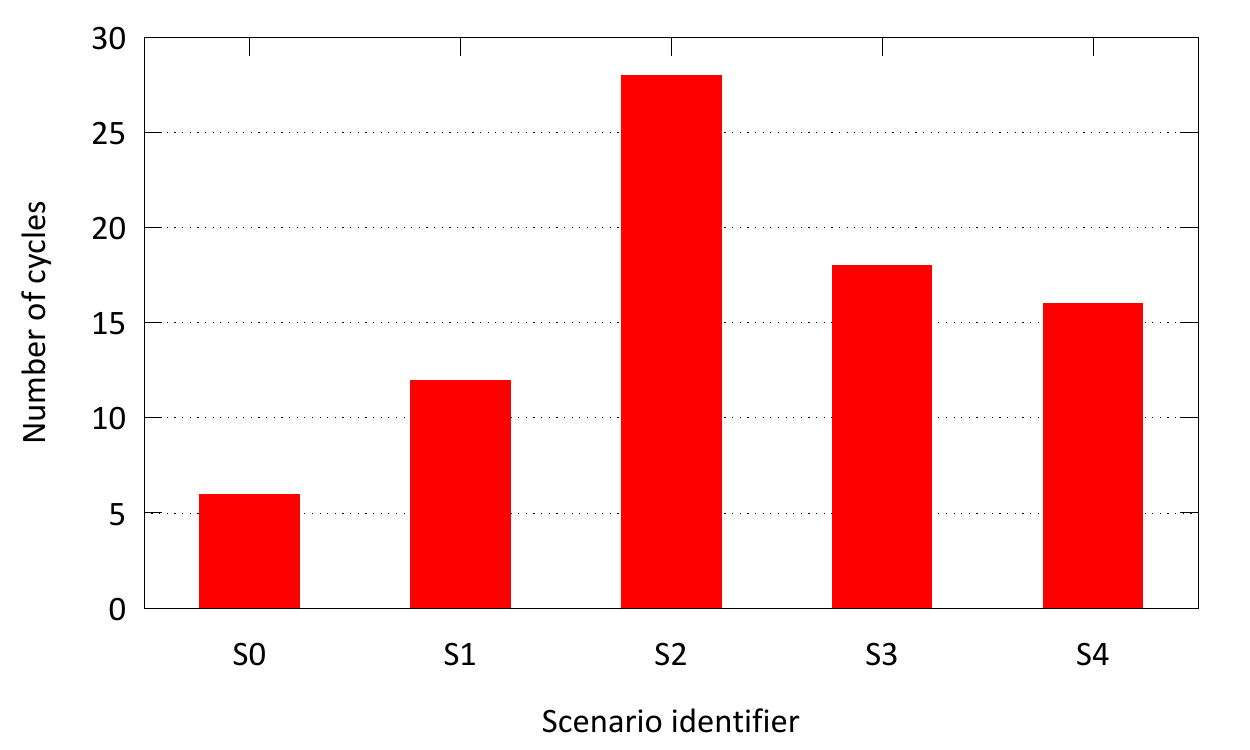}
		\caption{Latency results of scenarios}
		\label{fig18}
	\end{figure}

	In this section we propose to analyze the latency of realistic scenarios based on applications described in Section \ref{subsec_set} (\emph{picProc}, \emph{picDRM} and \emph{picDec}). Simulations were done with the following scenarios:
	\begin{itemize}
		\item S0: latency of a single \emph{Local Firewall}.
		\item S1: communication between two elements within the architecture (use of two \emph{Local Firewalls}).
		\item S2: communication between a processor and an external memory section protected with confidentiality and integrity (implies one \emph{Local Firewall} and a \emph{Cryptographic Firewall}). 
		\item S3: communication between a processor and an integrity only section in the external memory.
		\item S4: communication between a processor and a plaintext section in the external memory.
	\end{itemize}
	Latency of a \emph{Local Firewall} (scenario S0) serves as a reference, other scenarios imply two firewalls (two \emph{Local} for S1 while S2, S3 and S4 use one \emph{Local} and one \emph{Cryptographic Firewall}). Results are given in Figure \ref{fig18}.
	In terms of latency, the most critical scenario is S2 as it implies cryptographic operations (access to the external memory). S2 lasts for 28 clock cycles:
	\begin{itemize}
		\item \textbf{6 cycles} for the security policy checking as in a \emph{Local Firewall}, identically to scenario S0.
		\item According to AES-GCM, protection with confidentiality/integrity of N data blocks is performed in $10+(10+2) \times N$ cycles (therefore, \textbf{22 cycles} for a single data block).
	\end{itemize}
	Case study with 8KB caches (both instruction and data) is considered in this study. It is assumed that the overall execution time is divided in two parts:
	\begin{itemize}
		\item Accesses to the external memory (cache miss or read/write operations). It implies 1 \emph{Cryptographic Firewall} and 1 \emph{Local Firewall}.
		\item Internal accesses (between a processor and an IP or between two processors). This option only implies \emph{Local Firewalls}.
	\end{itemize}
	Timers and processor specific registers are used to extract information and compare latency overheads.
	\begin{table}[htbp]
		\centering
		\caption{Latency overheads for custom applications}
		\label{tab8}
		\begin{tabular}{c|c|c|c|}
			\cline{2-4}
			& {\bf \begin{tabular}[c]{@{}c@{}}Execution\\ time (ms)\end{tabular}} & {\bf \begin{tabular}[c]{@{}c@{}}\# of accesses\\ to external memory\end{tabular}} & {\bf \begin{tabular}[c]{@{}c@{}}Latency overhead\\ \cite{Fiorin08b} and this work\\ approach\end{tabular}} \\ \hline
			\multicolumn{1}{|c|}{{\bf picProc}} & 3,623 & 34,063,298 & 17.76\% \\ \hline
			\multicolumn{1}{|c|}{{\bf picDrm}} & 1,084 & 9,642,055 & 9.43\% \\ \hline
			\multicolumn{1}{|c|}{{\bf picDev}} & 382 & 4,736,966 & 4.18\% \\ \hline
		\end{tabular}
	\end{table}
	The simplest application (\emph{picDec}) has an overhead lower than 5\% (it does not require cryptographic features). \emph{picProc} is the application with the highest penalty (nearly 18\%) as it implies a \emph{Cryptographic Firewall} performing AES-GCM operations. In order to compare this work with existing solutions, firewalls approach is transposed in a centralized approach as it is proposed in the SECA model \cite{Coburn05}.
	\begin{figure}[htbp]
		\centering
		\subfloat[Distributed]{\label{fig19a}
			\includegraphics[width=.45\textwidth]{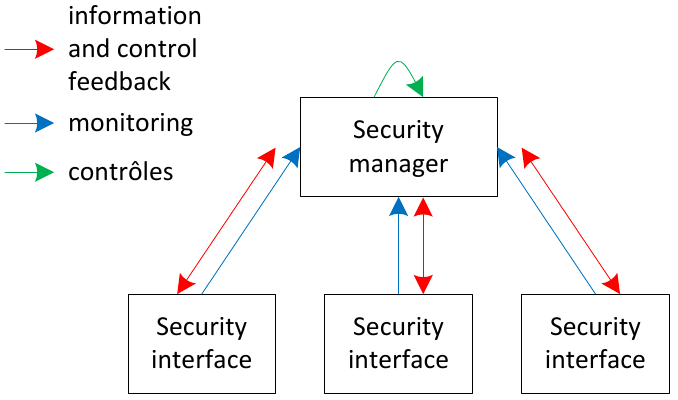}
		}
		\subfloat[Centralized]{\label{fig19b}
			\includegraphics[width=.45\textwidth]{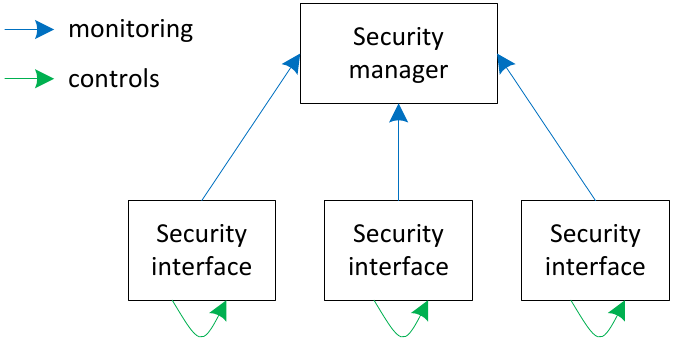}
		}
		\caption{Distributed and centralized behaviors}
		\label{fig19}
	\end{figure}
	~\\
	In a distributed approach (as firewalls, see Figure \ref{fig19a}), security controls are performed locally and monitoring information is sent to the security manager when an attack is detected (this is done by the monitoring IP and the update processor). In a centralized approach (\cite{Coburn05}, see Figure \ref{fig19b}), each security interface sends information to the security manager doing all the security controls. These connections are considered as trusted areas, but centralization implies latency overhead for communications between each interface and the manager. This work considers that each transaction (a roundtrip between an interface and a manager) through an AXI-Lite bus requires 4 additional clock cycles \cite{Arm12,Chang09}. Therefore, for the previously defined \emph{picDec} application, a centralized implementation gives a 6.18\% latency overhead while the distributed firewall approach has a 4.18\% overhead (a 33\% gain compared to SECA approach).
	
	\subsubsection{Memory occupancy}
	According to firewall description detailed in Section \ref{sec_static}, internal memories can be used to store:
	\begin{itemize}
		\item Security policies: parameters to be verified by firewalls.
		\item Timestamps: used by the AES-GCM core to protect against replay attacks. It is assumed that timestamps are generated with internal counters.
		\item Tags: when an external memory section has to be protected in terms of integrity, a tag is needed.
	\end{itemize}
	This work only focuses on tags produced by the AES-GCM core. For each 128-bit data, a tag of equal length is stored. Therefore, we can only protect 1.87MB of data in terms of integrity in a Xilinx ML605 board (XC6VLX240T1156-1 FPGA).
	\begin{itemize}
		\item At first sight, this can be a limiting factor. A more efficient approach would be the use of Bloom filters \cite{Bloom70,Crenne11b}: under specific conditions, this technique can decrease the memory footprint by 96\%.
		\item However, firewalls offer flexibility in terms of system security. In accordance with user requirements, data may not always be protected in terms of integrity (firewalls are able to detect other threats).
	\end{itemize}
	
	\section{Conclusion and future work}%
	Several related works show an interest in communication and memory protection. Nevertheless, no approach has addressed the need for cryptographic functions aiming at protecting memories with monitoring and controls function providing communication protection. The distributed firewall approach allows embedded system developers to protect both communications and memories in a multiprocessor architecture.
	
	Firewalls implementation shows that security can be provided only with hardware firewalls. The additional software is only for update purposes (attack detection, new configurations computation\ldots). This work proposes a low-latency solution where latency overheads can reach 17\% (according to cryptographic features set in firewalls).
	
	One perspective regarding our firewalls approach is a software protection: firewalls would analyze data according to the current software task identifier, which allows refining the security level of the overall solution. Finally, even if hardware security is a recent domain, it is possible to implement efficient functions; we believe that technology improvements (for instance, multi-layer FPGAs or hybrid platforms embedding FPGAs and processors) will allow going further in protection of multiprocessor architectures.
\bibliographystyle{ieeetr}
\bibliography{biblio-arxiv}
\end{document}